# A Logical Study of Partial Entailment

**Yi Zhou**                                                                    YZHOU@SCM.UWS.EDU.AU
**Yan Zhang**                                                                  YAN@SCM.UWS.EDU.AU
*Intelligent Systems Laboratory*
*School of Computing and Mathematics*
*University of Western Sydney, NSW, Australia*

## Abstract

We introduce a novel logical notion–partial entailment–to propositional logic. In contrast with classical entailment, that a formula $P$ *partially entails* another formula $Q$ with respect to a background formula set $\Gamma$ intuitively means that under the circumstance of $\Gamma$, if $P$ is true then some "part" of $Q$ will also be true. We distinguish three different kinds of partial entailments and formalize them by using an extended notion of prime implicant. We study their semantic properties, which show that, surprisingly, partial entailments fail for many simple inference rules. Then, we study the related computational properties, which indicate that partial entailments are relatively difficult to be computed. Finally, we consider a potential application of partial entailments in reasoning about rational agents.

## 1. Introduction

In standard propositional logic, classical entailment does not distinguish among the formulas that do not entail another formula. For example, let $x$, $y$ and $z$ be three atoms. We have that neither $x$ nor $z$ classically entails $x \wedge y$. However, $x$ seems intuitively "closer" to $x \wedge y$ than $z$. The reason is that, $x \wedge y$ can be considered as the union of two "parts" $x$ and $y$, and $x$ is exactly one of them. On the other hand, $z$ is completely irrelevant to $x \wedge y$.

This example motivates us to consider the notion of "partial entailment". In comparison with classical entailment, partial entailment intends to capture the "partial satisfaction" relationship between two formulas with respect to a background theory. In standard propositional logic, that a formula $P$ entails another formula $Q$ with respect to a formula set $\Gamma$ intuitively means that under the circumstance of $\Gamma$, if $P$ is true then $Q$ will also be true. In contrast, that a formula $P$ partially entails another formula $Q$ with respect to a formula set $\Gamma$ intuitively means that, under the circumstance of $\Gamma$, if $P$ is true then some "part" of $Q$ will also be true. Partial entailment is widely used in everyday life. For example, suppose that a student Laura wants to get score "HD" on the subjects mathematics and physics in the final examination. If she cannot achieve both of them, she could also be satisfied to achieve just one of them. At least, it is better than achieving none.

Let us consider a further example. Suppose that the background formula set is empty and the objective formula is $x \wedge y$. Consider the following four formulas: $x$, $x \wedge z$, $x \wedge \neg y$ and $z$. Clearly, none of these formulas classically entails $x \wedge y$. However, $x$, $x \wedge z$ and $x \wedge \neg y$ entail $x$, which can be regarded as a part of $x \wedge y$. In contrast, $z$ is irrelevant to $x \wedge y$. Thus, one can conclude that, to some extent, $x$, $x \wedge z$ and $x \wedge \neg y$ "partially satisfy" $x \wedge y$ while $z$ does not.

One may observe that there exists some difference between $x$ and $x \wedge z$ in partially satisfying $x \wedge y$. That is, $x \wedge z$ contains a new atom $z$, which is irrelevant to $x \wedge y$. On the other hand, $x$ is exactly a part of $x \wedge y$. In other words, $x \wedge z$ contains "irrelevancy" (namely $z$) in partially satisfying $x \wedge y$ while $x$ does not. Another observation is about $x \wedge \neg y$. Although it entails $x$ (a part of $x \wedge y$), it also entails $\neg y$, which contradicts to $y$ (also a part of $x \wedge y$). In other words, $x \wedge \neg y$ has some "side effect" (namely $\neg y$) besides partially satisfying $x \wedge y$.

The background theory is crucial for partial entailment. As an example, consider $z$ and $x \wedge y$ again. Suppose that the background formula set is empty. In this case, $z$ does not partially satisfy $x \wedge y$. If the background formula set turns into $\{z \rightarrow x\}$, then $z$ should partially satisfy $x \wedge y$ since under the background theory, if $z$ holds, then $x$ holds as well.

Based on the above observations and intuitions, in this paper, we formalize the notion of partial entailment between two formulas with respect to a background theory in a propositional language. Moreover, we distinguish three different kinds of partial entailments, namely *weak partial entailment*, *partial entailment* and *strong partial entailment*. The intuitions of them are summarized in Table 1. Of course, all of them require partial satisfaction between the two formulas. However, weak partial entailment may allow both irrelevancies and side effects, whilst partial entailment prohibits side effects but still may allow irrelevancies, and strong partial entailment is the strongest one that further prohibits irrelevancies.

Table 1: Intuitions of partial entailments

| - | partial satisfaction | irrelevancy | side effect |
|---|---|---|---|
| weak partial entailment | yes | allowed | allowed |
| partial entailment | yes | allowed | no |
| strong partial entailment | yes | no | no |

This paper is organized as follows. In the next section, we formalize three different kinds of partial entailments by using an extended notion of prime implicant, and discuss their semantic properties. In Section 3, we focus on the computational complexities of decision problems in relation to prime implicant and partial entailments, ranging from some special cases to the most general case. We then compare the notion of partial entailments with related notions in the AI literature in Section 4. In Section 5, we show how the notions of partial entailments can be applied to formalize partial goal satisfaction in reasoning about rational agents. Finally, we draw our conclusions in Section 6.

## 2. Partial Entailment

We restrict our discussions within a propositional language, denoted by $\mathcal{L}$. Formulas in $\mathcal{L}$ are composed recursively by a finite set *Atom* of *atoms* (also called *variables*) with $\{\top, \bot\}$ and standard connectives $\neg$ and $\rightarrow$. The connectives $\wedge$, $\vee$, $\leftrightarrow$ are defined as usual. *Literals* are atoms and their negations. We use lower case letters to denote atoms and literals, upper case letters to denote formulas, lower Greek letters to denote literal sets, and upper Greek letters to denote formula sets respectively. We write $-l$ to denote the complementary literal of a literal $l$, $-\pi$ to denote the set of complementary literals of all literals in $\pi$. We write $Atom(l)$, $Atom(P)$, $Atom(\pi)$ and $Atom(\Gamma)$ to denote the sets of atoms occurring in literal $l$, formula $P$, literal set $\pi$ and formula set $\Gamma$ respectively.

We say that a literal set $\pi$ is an *assignment* over a set $A$ of atoms (assignment for short if $A = Atom$) if for each atom $x \in A$, exactly one of $x$ and $\neg x$ is in $\pi$. Notice that both a literal $l$ and an assignment $\pi$ can also be considered as formulas. For convenience, henceforth, both $l$ and $\pi$ also denote their corresponding formulas if it is clear from the context. The *entailment relation* $\models$ and the notion of *model* are defined in the standard way. A set of formulas is said to be *consistent* if it has at least one model, otherwise, it is said to be *inconsistent*. A *theory* is a set of formulas closed under $\models$. Let $\Gamma$ be a set of formulas. The deductive closure of $\Gamma$, denoted by $Th(\Gamma)$, is the minimal theory (in the sense of set inclusion) containing $\Gamma$. For convenience, we also use $\Gamma$ itself to denote the theory $Th(\Gamma)$ if it is clear from the context.

We write $P|l$ to denote the formula obtained from $P$ by simultaneously replacing every occurrence of $x$ by $\top$ ($\bot$) if $l$ is of the form $x$ ($\neg x$). If $\pi$ is a consistent literal set (i.e. $\nexists x \in Atom$ s.t. $x \in \pi$ and $\neg x \in \pi$), and $\pi = \{l_1, l_2, ..., l_k\}$, we write $P|\pi$ to denote the formula $(((P|l_1)|l_2)|...)|l_k$.

## 2.1 A Simple Case

We begin to define the notions of partial entailments between two formulas with respect to a background formula set by considering a rather simple case. That is, both the two formulas are consistent conjunctions of literals and the background formula set is assumed to be empty.

**Definition 1** *Let $\pi$ and $\pi'$ be two consistent sets of literals.*

- *We say that $\pi$ weakly partially entails $\pi'$ if $\pi \cap \pi' \neq \emptyset$.*

- *We say that $\pi$ partially entails $\pi'$ if $\pi \cap \pi' \neq \emptyset$ and $\pi \cap -\pi' = \emptyset$.*

- *We say that $\pi$ strongly partially entails $\pi'$ if $\emptyset \subset \pi \subseteq \pi'$.*

Definition 1 simply follows the intuitions presented in the introduction section (see Table 1). In this simple case, $\pi'$ is a consistent set of literals. Thus, the "parts" of $\pi'$ can be regarded as all the elements (or subsets) of $\pi'$. Recall our intuitive sense of partial satisfaction. That is, if $\pi$ holds, then some parts of $\pi'$ hold as well. This means that there exists some part of $\pi'$, which is a subset of $\pi$. Clearly, this can be precisely captured by $\pi \cap \pi' \neq \emptyset$, which is exactly the definition of weak partial entailment in this simple case.

For partial entailment, we forbid side effects based on partial satisfaction. Again, since $\pi'$ is a consistent set of literals, the "side effects" of $\pi'$ can be regarded as all the complementary literal of the elements (or subsets) of $\pi'$. Hence, that $\pi$ has no side effects to $\pi'$ means that $\pi$ does not mention the complementary literal of any elements (or subsets) of $\pi'$. Clearly, this can be precisely captured by $\pi \cap -\pi' = \emptyset$, where $-\pi'$ is the set of complementary literals of all elements in $\pi'$. Together with $\pi \cap \pi' \neq \emptyset$, this forms the definition of partial entailment in this simple case.

Finally, consider strong partial entailment. We require additionally that there is no irrelevancy. The irrelevancies of $\pi'$ can be considered as all those atoms (or literals) not mentioned in $\pi'$. Hence, that $\pi$ has no irrelevancy to $\pi'$ means that there is no atom (or literal) mentioned in $\pi$ but not in $\pi'$. Formally, this can be precisely captured by $Atom(\pi) \subseteq Atom(\pi')$ (or $\pi \subseteq \pi'$). Whatever the case is, together with the two conditions $\pi \cap \pi' \neq \emptyset$ and $\pi \cap -\pi' = \emptyset$, this is equivalent to $\pi \subseteq \pi'$ and $\pi \neq \emptyset$, which is the definition of strong partial entailment in this simple case.

**Example 1** *Recall the example proposed in the introduction section. According to Definition 1, it is easy to check that $x$, $x \wedge \neg y$ and $x \wedge z$ weakly partially entail $x \wedge y$ but $z$ does not; $x$ and $x \wedge z$ partially entail $x \wedge y$ but $x \wedge \neg y$ and $z$ do not; only $x$ strongly partially entails $x \wedge y$ while $x \wedge \neg y$, $x \wedge z$ and $z$ do not. The results coincide with our intuitions and observations discussed in the introduction section.*

*Comparing weak partial entailment and partial entailment, one may observe that $x \wedge \neg y$ weakly partially entails $x \wedge y$ but it does not partially entail $x \wedge y$. As discussed, this is because $\neg y$ is a "side effect" to $x \wedge y$. Comparing partial entailment and strong partial entailment, one may observe that $x \wedge z$ partially entails $x \wedge y$ but it does not strongly partially entail $x \wedge y$. Again, this coincides with our discussions since $z$ is an "irrelevancy" to $x \wedge y$. $\square$*

An interesting phenomenon is about the relationships between partial entailments and classical entailment. In this simple case, it is easy to see that if $\pi$ classically entails $\pi'$, then $\pi$ partially entails $\pi'$. Also, $\pi$ weakly partially entails $\pi'$. This means that classical entailment is a special case of (weak) partial entailment to some extent. However, this does not hold for strong partial entailment. For instance, $x \wedge y$ classically entails $x$, but it does not strongly partially entail $x$ since $y$ is an "irrelevancy" with respect to $x$.

At first glance, it seems that this result is strange in the sense that the term of partial entailment should be a generalization of classical entailment. However, this does not mean that classical entailment also prohibits side effects and irrelevancies. Thus, from the intuitive sense, one can only conclude that classical entailment should be a special case of weak partial entailment, but may be mutually different from partial entailment and strong partial entailment. Interestingly, as we will show later, classical entailment indeed prohibits side effects. This means that classical entailment is a special case of partial entailment as well. However, classical entailment may allow irrelevancy (e.g., $x \wedge y$ classically entails $x$ but $y$ is an irrelevancy to $x$). Hence, roughly speaking, classical entailment is a special case of both partial entailment and weak partial entailment, but classical entailment and strong partial entailment are mutually different.

Definition 1 gives a basic impression on the notions of partial entailments. In the following, we consider to extend the three notions of partial entailments in a general sense, in which the two formulas are arbitrary formulas and the background is an arbitrary set of formulas.

## 2.2 Prime Implicant

In order to define partial entailments in general case, we use Quine's notion of prime implicant (Quine, 1952), which has been widely used in many areas in logic and computer science. An excellent survey of prime implicant and its dual prime implicate is due to Marquis (2000).

Roughly speaking, a prime implicant of a formula is a minimal set (in the sense of set inclusion) of literals that entails this formula. The notion of prime implicant can be relativized with respect to a background formula set as follows.

**Definition 2 (Relativized prime implicant)** *A literal set $\pi$ is a* prime implicant *of a formula $P$ with respect to a formula set $\Gamma$ if:*

1. *$\Gamma \cup \pi$ is consistent.*

2. *$\Gamma \cup \pi \models P$.*

3. *There does not exist a literal set $\pi' \subset \pi$ that satisfies the above two conditions.*

*The set of all prime implicants of $P$ with respect to $\Gamma$ is denoted by $PI(\Gamma, P)$. For convenience, we omit $\Gamma$ when it is empty.*

Intuitively, a prime implicant of $P$ *w.r.t.* $\Gamma$ is a minimal set of information needed in order to satisfy $P$ under the circumstance of $\Gamma$. Condition 1 requires that the set of information is "feasible", i.e., it must be consistent with the background formula set. Condition 2 means that the set of information is an "implicant", i.e., it is indeed powerful enough to satisfy $P$ *w.r.t.* $\Gamma$. Finally, condition 3 requires that the set of information is "prime", i.e., it is a minimal set of information for satisfying $P$ *w.r.t.* $\Gamma$.

**Example 2** *According to Definition 2, the only prime implicant of $x \wedge y$ is $\{x, y\}$, while $x \vee y$ has two prime implicants, namely $\{x\}$ and $\{y\}$. The background formula set indeed plays an important role. Clearly, $x$ has a unique prime implicant $\{x\}$. However, $\{y\}$ is also a prime implicant of $x$ with respect to the background formula set $\{y \rightarrow x\}$.* □

It is well known that the relativized notion of prime implicant is similar to the notion of logic-based abduction (Eiter & Gottlob, 1995; Selman & Levesque, 1990; Eiter & Makino, 2007). There are different characterizations of abduction in the literature. Here, we consider one of the most well-studied forms (Selman & Levesque, 1990; Eiter & Makino, 2007).

**Definition 3 (Selman & Levesque, 1990)** *Let $\Gamma$ be a theory, $F$ a formula and $H$ a set of literals. An* (abductive) explanation *of $F$ with respect to $H$ from $\Gamma$ is a minimal set $\pi \subseteq H$ such that:*

1. *$\Gamma \cup \pi$ is consistent,*

2. *$\Gamma \cup \pi \models F$.*

Following the above definitions, it can be observed that the relativized notion of prime implicant and abduction can be simply transformed each other. More precisely, suppose that $\Gamma$ is a theory, $F$ a formula and $H$ a set of literals. Let *Lit* be the set of all literals in the language. Then, $\pi$ is a prime implicant of $F$ *w.r.t.* $\Gamma$ iff $\pi$ is an abductive explanation of $F$ from $\Gamma$ with respect to *Lit*. Conversely, $\pi$ is an abductive explanation of $F$ with respect to $H$ from $\Gamma$ iff $\pi$ is a prime implicant of $F$ *w.r.t.* $\Gamma$ and mentions literals only from $H$.

Another closely related notion to prime implicant is its dual, called *prime implicate*, which is also of interests in many areas as well. Roughly speaking, a prime implicate of a propositional formula $P$ is a minimal clause (i.e., disjunction of literals) entailed by $P$. It is easy to see that the conjunction of a set $\pi$ of literals is a prime implicant of a formula $P$ if and only if the disjunction of all the complementary literals of the elements in $\pi$ is a prime implicate of $\neg P$. Prime implicate can also be relativized to a background formula set as well (Marquis, 2000). We omit this definition since we are mainly focused on prime implicant in this paper.

Many properties in relation to prime implicate were discussed and summarized by Marquis (2000). Not surprisingly, similar results hold for prime implicant as well. In the following, we recall some of the important properties in relation to the relativized notion of prime implicant, which will be used later in this paper.

**Proposition 1** *Let $\Gamma$ be a finite set of formulas, $P$ a formula and $\pi$ a set of literals. $\pi$ is a prime implicant of $P$ w.r.t. $\Gamma$ iff $\Gamma \cup \pi$ is consistent and $\pi$ is a prime implicant of $\bigwedge \Gamma \to P$, where $\bigwedge \Gamma$ is the conjunction of all formulas in $\Gamma$.*

**Proposition 2** *Let $P$ be a formula, $\Gamma$ a set of formulas and $\pi$ a literal set such that $\Gamma \cup \pi$ is consistent and $\Gamma \cup \pi \models P$. Then there exists $\pi' \subseteq \pi$ such that $\pi'$ is a prime implicant of $P$ with respect to $\Gamma$.*

**Proposition 3** *Let $P$ be a formula, $\Gamma$ a formula set and $\pi$ a literal set. If $\pi$ is a prime implicant of $P$ w.r.t. $\Gamma$, then there exists an assignment $\pi'$ such that $\pi \subseteq \pi'$ and $\pi'$ is a model of both $\Gamma$ and $P$.*

Together with Proposition 2, Proposition 3 indicates the correspondence relationship between all the models of $P$ w.r.t. $\Gamma$ and all the prime implicants of $P$ w.r.t. $\Gamma$. Suppose that $\pi$ is a model of $P$ and is consistent with $\Gamma$. Then, according to Proposition 2, there exists a subset of $\pi$, which is a prime implicant of $P$ w.r.t. $\Gamma$. Conversely, according to Proposition 3, every prime implicant of $P$ w.r.t. $\Gamma$ can be extended to a model of both $\Gamma$ and $P$.

**Proposition 4** *Let $P$ be a formula and $\Gamma$ a set of formulas. $\Gamma \models \neg P$ if and only if $PI(\Gamma, P) = \emptyset$; $\Gamma \models P$ if and only if $PI(\Gamma, P) = \{\emptyset\}$.*

**Proposition 5** *Let $P$ and $Q$ be two formulas and $\Gamma$ a set of formulas. $\Gamma \models P \leftrightarrow Q$ iff $PI(\Gamma, P) = PI(\Gamma, Q)$.*

To conclude, the prime implicants of a formula $P$ w.r.t. a formula set $\Gamma$ play two roles.

*Cases*: On one hand, if there exists an assignment $\pi$ satisfying both $\Gamma$ and $P$, then according to Proposition 2, there exists a subset of $\pi$ which is a prime implicant of $P$ w.r.t. $\Gamma$. On the other hand, if $\pi$ is a prime implicant of $P$ w.r.t. $\Gamma$, then according to Proposition 3, it can be extended to an assignment satisfying both $\Gamma$ and $P$. This means that the prime implicants of $P$ w.r.t. $\Gamma$ are corresponding to the possible worlds (assignments) satisfying both $\Gamma$ and $P$. Intuitively, they are corresponding to all the *cases* which make $P$ true w.r.t. $\Gamma$, i.e., all the ways to achieve $P$ under $\Gamma$.

*Parts*: Suppose that $\pi$ is a prime implicant of $P$ *w.r.t.* $\Gamma$ and $l$ is a literal in $\pi$. Then, $\Gamma \cup \pi \models P$ and $\Gamma \cup \pi \backslash \{l\} \not\models P$. As mentioned above, $\pi$ can be considered as a way (case) to achieve $P$ *w.r.t.* $\Gamma$. Then, intuitively, $l$ plays an essential role for achieving $P$ *w.r.t.* $\Gamma$ via $\pi$. Without $l$, $\pi$ is no longer a way to achieve $P$ under $\Gamma$. This shows that all the elements in $\pi$ are essential. In other words, $\pi$ is a minimal way to achieve $P$ under $\Gamma$, and none of the elements in $\pi$ can be thrown away. Thus, these literals can be considered as *parts* of $P$ *w.r.t.* $\Gamma$.

Both roles of prime implicants will be exploited by the notions of partial entailments introduced in the following.

## 2.3 Definitions of Partial Entailments

Based on the notion of prime implicant, we formalize three kinds of partial entailments. From the weakest to the strongest, they are weak partial entailment, partial entailment and strong partial entailment respectively.

**Definition 4 (Weak partial entailment)** *A formula $P$ weakly partially entails a formula $Q$ with respect to a formula set $\Gamma$, denoted by $P \succ_\Gamma^W Q$, if:*

1. *$PI(\Gamma, P)$ is not empty.*

2. *For each $\pi \in PI(\Gamma, P)$, there exists $\pi' \in PI(\Gamma, Q)$, such that $\pi \cap \pi' \neq \emptyset$.*

**Definition 5 (Partial entailment)** *A formula $P$ partially entails a formula $Q$ with respect to a formula set $\Gamma$, denoted by $P \succ_\Gamma Q$, if:*

1. *$PI(\Gamma, P)$ is not empty.*

2. *For each $\pi \in PI(\Gamma, P)$, there exists $\pi' \in PI(\Gamma, Q)$, such that $\pi \cap \pi' \neq \emptyset$ and $\pi \cap -\pi' = \emptyset$.*

**Definition 6 (Strong partial entailment)** *A formula $P$ strongly partially entails a formula $Q$ with respect to a formula set $\Gamma$, denoted by $P \succ_\Gamma^S Q$, if:*

1. *$PI(\Gamma, P)$ is not empty.*

2. *For each $\pi \in PI(\Gamma, P)$, there exists $\pi' \in PI(\Gamma, Q)$, such that $\emptyset \subset \pi \subseteq \pi'$.*

We write $P \not\succ_\Gamma^W Q$ if it is not the case that $P \succ_\Gamma^W Q$, similar for partial entailment and strong partial entailment. For convenience, we omit $\Gamma$ when it is empty.

Clearly, these definitions are generalizations of Definition 1 proposed in Section 2.1 since the unique prime implicant of a literal set is itself. Note that the only difference among these definitions of partial entailments is indeed presented in Section 2.1 in the simple case. Definitions 4-6 all require that for any literal set in $PI(\Gamma, P)$, there exists some literal set in $PI(\Gamma, Q)$ such that the two literal sets satisfy the same corresponding relationships presented in Definition 1 respectively.

Let us take a closer look at condition 2 in Definition 4. Recall our intuitive sense of partial satisfaction again, that is, if $P$ is true under $\Gamma$, then some parts of $Q$ hold as well.

As we discussed in Section 2.2, the prime implicants of $P$ *w.r.t.* $\Gamma$ represent all the cases that make $P$ true under $\Gamma$. Also, the prime implicants of $Q$ *w.r.t.* $\Gamma$ can capture the idea of "part" of $Q$ under $\Gamma$. Hence, condition 2 means that, intuitively, for all the cases that $P$ is true under $\Gamma$ (captured by all the prime implicants of $P$ *w.r.t.* $\Gamma$), there exists a way to achieve $Q$ under $\Gamma$ (captured by a prime implicant of $Q$ *w.r.t.* $\Gamma$), such that the former partially satisfies the latter (reduced to the simple case in Definition 1). In addition, condition 1 in Definition 4 ensures that there exists at least one such situation. The intuitive senses of Definitions 5 and 6 can be explained in a similar way.

**Example 3** *Let the background theory* $\Gamma$ *be* $\{x \vee y, z \to y\}$. *Consider two formulas* $P = (x \wedge r) \vee (y \wedge s)$ *and* $Q = (x \wedge z) \vee (\neg x \wedge y \wedge s)$. *The prime implicants of* $P$ *w.r.t.* $\Gamma$ *are as follows:* $\{x, r\}$, $\{y, s\}$, $\{r, s\}$, $\{z, s\}$, $\{\neg y, r\}$ *and* $\{\neg x, s\}$. *Notice that* $\{\neg y, r\}$ *is a prime implicant of* $P$ *w.r.t.* $\Gamma$ *although the literal* $\neg y$ *even does not occur in P. On the other hand, the prime implicants of* $Q$ *w.r.t.* $\Gamma$ *are as follows:* $\{x, z\}$, $\{\neg x, s\}$ *and* $\{z, s\}$. *Notice that* $\{\neg x, y, s\}$ *is not a prime implicant of* $Q$ *w.r.t.* $\Gamma$ *since* $\{\neg x, s\}$ *is a subset of it and also satisfies* $Q$ *under* $\Gamma$.

*According to the definitions, we have that* $P$ *does not weakly partially entail* $Q$ *w.r.t.* $\Gamma$ *since there is a prime implicant* $\{\neg y, r\}$ *of* $P$ *w.r.t.* $\Gamma$ *such that it has no intersection with any of the prime implicants of* $Q$ *w.r.t.* $\Gamma$. *Also,* $P$ *does not (strongly) partially entail* $Q$ *w.r.t.* $\Gamma$ *either. Conversely,* $Q$ *(weakly) partially entails* $P$ *w.r.t.* $\Gamma$, *whilst it does not strongly partially entail* $P$ *w.r.t.* $\Gamma$. $\square$

Note that there are other possible relations between two formulas *w.r.t.* a background formula set by using the extended notion of prime implicant. In this paper, for defining partial entailments, we use a $\forall\exists$ style definition in the sense that we require that "for all" literal sets in $PI(\Gamma, P)$, "there exists" a literal set in $PI(\Gamma, Q)$ such that the two literal sets satisfy the conditions in Definition 1 respectively. Recall again the intuition of partial entailments, which requires that in any cases, if $P$ is true under the context of $\Gamma$, then some part of $Q$ is true under the context of $\Gamma$. This naturally suggests the $\forall\exists$ style definition for the definitions of partial entailments (i.e. Definitions 4-6), where the "$\forall$"-part captures all the cases of achieving $P$ under the context of $\Gamma$, and the "$\exists$"-part associates a possibility of achieving $Q$ under the context of $\Gamma$ to the above case for $P$. In addition, the condition item 2 in Definitions 4-6 ensures that, in this case, $P$ partially achieves $Q$ *w.r.t.* $\Gamma$ via the two literal sets.

The reason why we choose the $\forall\exists$ style definition for partial entailments can also be explained analogously to classical entailment. Note that a formula $P$ classically entails another formula $Q$ *w.r.t.* a background theory $\Gamma$ iff for all models of $P$ *w.r.t.* $\Gamma$, it is also a model of $Q$ *w.r.t.* $\Gamma$.

Certainly, there are other possible styles, including $\forall\forall$, $\exists\exists$ and $\exists\forall$. For instance, the $\exists\forall$ style definition in the weakest case should be: "there exists" $\pi \in PI(\Gamma, P)$ such that "for all" $\pi' \in PI(\Gamma, Q)$, $\pi \cap \pi' \neq \emptyset$. Although these definitions might be interesting and useful elsewhere, they fail to capture the basic idea of "partial satisfaction".

**Example 4** *Consider two formulas* $x \vee z$ *and* $x \wedge y$. *The former has two prime implicants, namely* $\{x\}$ *and* $\{z\}$, *while the latter has a unique prime implicant, namely* $\{x, y\}$. *If we use the* $\exists\exists$ *style definition, we have* $x \vee z$ *(weakly, strongly) partially entails* $x \wedge y$. *However,* this is not intuitive because $\{z\}$ achieves $x \vee z$ true but it is irrelevant to $x \wedge y$. In other words, the assignment $\{\neg x, \neg y, z\}$ satisfies $x \vee z$ but not $x \wedge y$. Hence, there exists a case that $x \vee z$ is true but none of the parts of $x \wedge y$ is true.

Consider the formula $x \vee y$, which has two prime implicants, $\{x\}$ and $\{y\}$. If we take the $\forall\forall$ style definition or the $\exists\forall$ one, this formula does not (weakly, strongly) partially entail itself. This is obviously not intuitive. □

Another feasible direction of definitions is to switch the formulas $P$ and $Q$. For instance, the switched-$\forall\exists$ style definition in the weakest case is: "for all" $\pi \in PI(\Gamma, Q)$, "there exists" some $\pi' \in PI(\Gamma, P)$ such that $\pi \cap \pi' \neq \emptyset$. This definition, and the corresponding switched-$\exists\forall$, $\exists\exists$ and $\forall\forall$ style definitions, fail to capture the basic idea of partial satisfaction either. Again, recall the intuition of partial entailments. That is, if $P$ is true (under the context of $\Gamma$), then some part of $Q$ is true as well (under the context of $\Gamma$). Similar to classical entailment, this naturally suggests the order of $P$ and $Q$ for the definitions of partial entailments. Also, the switched-$\forall\forall$ (switched-$\exists\exists$) style definition is the same as the $\forall\forall$ ($\exists\exists$) style definition. As shown in Example 4, they are not suitable for defining the notions of partial entailments. To understand why the switched-$\forall\exists$ style and the switched-$\exists\forall$ style definitions fail for capturing partial entailments, let us consider the following example.

**Example 5** *Consider two formulas $x \leftrightarrow y$ and $x$. The former has two prime implicants, namely $\{x, y\}$ and $\{\neg x, \neg y\}$, while the latter has a unique prime implicant, namely $\{x\}$. If we take the switched-$\forall\exists$ style definition, we have $x \leftrightarrow y$ (weakly, strongly) partially entails $x$. However, this is not intuitive because $\{\neg x, \neg y\}$ achieves $x \leftrightarrow y$ but it also achieves $\neg x$. Hence, there exists a case that $x \leftrightarrow y$ is true but $x$ is false.*

*Again, consider the formula $x \vee y$. It has two prime implicants, namely $\{x\}$ and $\{y\}$. Intuitively, this formula should (weakly, strongly) partially entail itself. However, this is not the case if we apply the switched-$\exists\forall$ style definition.* □

One of the fundamental properties is the relationships among three notions of partial entailments. The following proposition shows that weak partial entailment is weaker than partial entailment, which is further weaker than strong partial entailment. However, by observations from Example 1, the converses do not hold in general.

**Proposition 6 (Basic relationships among partial entailments)** *Let $P, Q$ be two formulas and $\Gamma$ a formula set. If $P \succ_\Gamma^S Q$, then $P \succ_\Gamma Q$; if $P \succ_\Gamma Q$, then $P \succ_\Gamma^W Q$.*

It is worth mentioning that, in the definitions of partial entailments, $PI(\Gamma, P)$ is required not to be empty to exclude the case when $P$ is inconsistent with the background theory $\Gamma$. The underlying intuition is that partial entailments require "real" connections between the two formulas *w.r.t.* the background theory. Let us take a closer look at item 2 in the definitions. The basic idea of partial entailments is that for each prime implicant of $P$ *w.r.t.* $\Gamma$, there exists a prime implicant of $Q$ *w.r.t.* $\Gamma$ such that the two literal sets satisfy that, for instance, their intersection are not empty. In other words, $P$ partially entails $Q$ *w.r.t.* $\Gamma$ via these two literal sets. However, item 2 does not exclude the case where there is no prime implicant of $P$ *w.r.t.* $\Gamma$. In this case, item 2 still holds. However, we cannot say that $P$ partially entails $Q$ *w.r.t.* $\Gamma$ "via the two literal sets" now because the two literals sets do not exist at all. Hence, we require an additional item 1 to make sure that $PI(\Gamma, P)$ is not empty.

Then, for any formulas $P$ inconsistent with $\Gamma$, $P$ does not (weakly, strongly) partially entail any formula $Q$. Also, for any formulas $P$ entailed by $\Gamma$, $P$ does not (weakly, strongly) partially entail any formula $Q$. However, the reasons of these two cases are different. For the former, the reason is that $PI(\Gamma, P)$ is empty (see Proposition 4), thus $P$ is excluded by item 1 in the definitions. For the latter, the reason is that $PI(\Gamma, P) = \{\emptyset\}$ (see Proposition 4 again), thus the prime implicant of $P$ *w.r.t.* $\Gamma$ has no intersections with other prime implicants. We say that a formula $P$ is *trivial* with respect to a background formula set $\Gamma$ if $\Gamma \models P$ or $\Gamma \models \neg P$. Otherwise, we say that $P$ is *nontrivial* with respect to $\Gamma$. The following proposition shows that a trivial formula can neither (weakly, strongly) partially entail any formula, nor they can be (weakly, strongly) partially entailed by any formulas.

**Proposition 7 (Non-Triviality)** *Let $\Gamma$ be a formula set and $P$ and $Q$ two formulas. If $P$ is trivial w.r.t. $\Gamma$, then $P \not\succ_\Gamma Q$ and $Q \not\succ_\Gamma P$. This assertion holds for weak partial entailment and strong partial entailment as well.*

**Proposition 8 (Extension of Classical Entailment)** *Let $\Gamma$ be a formula set and $P$ and $Q$ two formulas nontrivial w.r.t. $\Gamma$. If $\Gamma \models P \to Q$, then $P \succ_\Gamma Q$. Also, $P \succ_\Gamma^W Q$.*

Proposition 8 shows that partial entailment is an extension of classical entailment if only considering nontrivial formulas, so is weak partial entailment. The converses do not hold. As a simple example, $x$ (weakly) partially entails $x \wedge y$ but $x$ does not classically entail $x \wedge y$.

Note that strong partial entailment and classical entailment are mutually different. For example, $x \wedge y$ classically entails $x$ but $x \wedge y$ does not strongly partially entail $x$. Conversely, $x$ strongly partially entails $x \wedge y$ but $x$ does not classically entail $x \wedge y$. As we demonstrated in Section 2.1, the reason is that strong partial entailment prohibits irrelevancy whilst classical entailment does not.

In addition, strong partial entailment captures the notion of "part" in a broader sense. In other words, if a formula $P$ strongly partially entails a formula $Q$ with respect to a background formula set $\Gamma$, then $P$ can be considered as a "part" of $Q$ *w.r.t.* $\Gamma$. In this sense, it also explains why classical entailment does not necessarily imply strong partial entailment. However, this explanation seems to suggest the other way around. That is, if $P$ strongly partially entails $Q$ *w.r.t.* $\Gamma$ (i.e., $P$ is a part of $Q$ *w.r.t.* $\Gamma$), then $Q$ classically entails $P$ *w.r.t.* $\Gamma$. In fact, this is not the case either. For example, $x$ strongly partially entails $(x \wedge y) \vee (y \wedge z)$, but $(x \wedge y) \vee (y \wedge z)$ does not classically entails $x$. The reason here is that a formula may contain disjunctive information. If restricted into the simple case discussed in Section 2.1, this is indeed the case. That is, given two consistent literal sets $\pi$ and $\pi'$, $\pi$ strongly partially entails $\pi'$ if and only if $\pi'$ classically entails $\pi$.

## 2.4 More Semantic Properties

In this subsection, we extensively study some semantic properties in relation to all three kinds of partial entailments. The reasons why we are interested in these properties are twofold. Firstly, they provide deeper understandings of how partial entailments work. As we will see later, surprisingly, many simple inference rules fail for partial entailments. Secondly, these properties illustrate some similarities/differences among three kinds of partial entailments, as well as the similarities/differences between partial entailments and classical entailment.

We consider a collection of inference rules. Some of them are considered to be important in many kinds of philosophical logics and knowledge representation logics, others are likely to hold for partial entailments. Let $\Gamma$ and $\Gamma'$ be two sets of formulas, $P$, $Q$ and $R$ formulas and $x$ and $y$ atoms. In this subsection, we assume that all formulas are nontrivial *w.r.t.* the background theory.[1] Here, we use $>_\Gamma$ to denote any kinds of partial entailments with respect to $\Gamma$. The inferences rules we considered are listed as follows:

**Ref:** (Reflexivity) $P >_\Gamma P$, meaning that a formula partially entails itself *w.r.t.* any background theory.

**LE:** (Left Equivalence) If $\Gamma \models P \leftrightarrow R$ and $P >_\Gamma Q$, then $R >_\Gamma Q$, meaning that if two formulas are equivalent under the background theory, then they play the same role in partial entailments on the left side.

**RE:** (Right Equivalence) If $\Gamma \models Q \leftrightarrow R$ and $P >_\Gamma Q$, then $P >_\Gamma R$, meaning that if two formulas are equivalent under the background theory, then they play the same role in partial entailments on the right side.

**BE:** (Background theory Equivalence) If $\Gamma \models \Gamma'$ and $\Gamma' \models \Gamma$, then $P >_\Gamma Q$ iff $P >_{\Gamma'} Q$, meaning that if two background theories are equivalent, then they play the same role in partial entailments.

**Rev:** (Relevancy) If $P > Q$, then $Atom(P) \cap Atom(Q) \neq \emptyset$, meaning that if one formula partially entails another *w.r.t.* the empty background theory, then the atom sets used in the two formulas respectively must not be disjoint.

**Tran:** (Transitivity) If $P >_\Gamma Q$ and $Q >_\Gamma R$, then $P >_\Gamma R$, meaning that partial entailment relation among formulas is an ordering on the propositional language.

**AS:** (Atom Substitution) If $P >_\Gamma Q$, then $P(x/y) >_\Gamma Q(x/y)$,[2] meaning that any atoms in a partial entailment relation can be replaced by other atoms.

**LO:** (Left Or) If $P >_\Gamma Q$ and $R >_\Gamma Q$, then $P \lor R >_\Gamma Q$, meaning that if both two formulas partially entail another one, then their disjunction should partially entails the formula as well.

**LS:** (Left Strengthening) If $\Gamma \models P \rightarrow R$ and $R >_\Gamma Q$, then $P >_\Gamma Q$, meaning that if a formula partially entails another, then a formula strengthened by the former should partially entail the latter as well.

**RA:** (Right And) If $P >_\Gamma Q$ and $P >_\Gamma R$, then $P >_\Gamma R \land Q$, meaning that if two formulas are both partially entailed by another formula, then their conjunction should be partially entailed by the formula as well.

---

1. This is because trivial formulas are not interesting for partial entailments (see Proposition 7).
2. Here, $P(x/y)$ is the formula obtained from $P$ by simultaneously replacing every occurrence of atom $x$ by $y$, similar for $Q(x/y)$.

Table 2: Properties of partial entailments

| - | weak partial entailment | partial entailment | strong partial entailment |
|---|---|---|---|
| Ref | yes | yes | yes |
| LE | yes | yes | yes |
| RE | yes | yes | yes |
| BE | yes | yes | yes |
| Rev | yes | yes | yes |
| * Tran | no | no | yes |
| As | no | no | no |
| LO | no | no | no |
| * LS | yes | no | no |
| RA | no | no | no |
| RO | no | no | no |
| Mono | no | no | no |
| LN | no | no | no |
| RN | no | no | no |

**RO:** (Right Or) If $P >_\Gamma Q$, then $P >_\Gamma Q \vee R$, meaning that if a formula partially entails another, then it should partially entail the disjunction of the latter formula with any formulas.

**Mono:** (Monotonicity) If $\Gamma' \models \Gamma$ and $P >_\Gamma Q$, then $P >_{\Gamma'} Q$, meaning that adding more information into the background theory preserves partial entailment relations.

**LN:** (Left Negation) If $P >_\Gamma Q$, then $\neg P \not>_\Gamma Q$, meaning that it is impossible that both a formula and its negation partially entail another formula.

**RN:** (Right Negation) If $P >_\Gamma Q$, then $P \not>_\Gamma \neg Q$, meaning that it is impossible that both a formula and its negation can be partially entailed by another formula.

Suppose that we define the relation $>_\Gamma$ between $P$ and $Q$ as classical entailment, i.e., $\Gamma \models P \rightarrow Q$. Then, it satisfies all the inference rules proposed above. However, the following proposition shows that partial entailments fail for many of them.

**Proposition 9** *Table 2 summarizes whether all three kinds of partial entailments satisfy the inference rules considered above.*

Table 2 illustrates some similarities and differences among all three kinds of partial entailments. They have the same results for most of these inference rules. However, Transitivity (Tran) and Left Strengthening (LS) (highlighted by the * symbol in Table 2) are two exceptions. Transitivity (Tran) holds for strong partial entailment but not for partial entailment and weak partial entailment. Left Strengthening (LS) holds for weak partial entailment but not for partial entailment and strong partial entailment. Finally, both weak partial entailment and partial entailment are extensions of classical entailment in the sense considered in Proposition 8, while strong partial entailment is not.

Also, Table 2 illustrates some similarities and differences between partial entailments and classical entailment. As we mentioned above, all these inference rules hold for classical entailment. However, this is not the case for partial entailments. This means that partial entailments actually behave quite different from classical entailment. Among these inference rules, Transitivity is an important one. It does not hold for partial entailment and weak partial entailment since they allow irrelevancies. Nevertheless, Transitivity holds for strong partial entailment. However, strong partial entailment is even more dissimilar to classical entailment since they are mutually different.

Table 2 also indicates that it is not easy to capture the properties of the notions of partial entailments due to the fact that they fail most of the simple inference rules (see Table 2). Meanwhile, some properties, (e.g. Tran, LN and RN), may seem to be intuitive for partial entailments. However, it turns out that they do not hold in general according to the formal definitions.

Certainly, one can consider other inference rules, for instance, Contraposition. That is, if $P >_\Gamma Q$, then $\neg Q \succ_\Gamma \neg P$. One can check that Contraposition does not hold for all three kinds of partial entailments either.

## 2.5 Discussions

One may doubt whether partial entailments can be directly captured within classical propositional logic. For example, one may define partial entailment as follows. A formula $P$ partially entails a formula $Q$ w.r.t. a formula set $\Gamma$ iff there exists a formula $R$ such that $\Gamma \cup \{R\} \not\models Q$ but $\Gamma \cup \{P, R\} \models Q$. However, this definition cannot capture partial satisfaction. In fact, for every formula $P$ such that $\Gamma \not\models P \to Q$, there always exists such a formula $R$ (Let $R$ be $\neg \bigwedge \Gamma \vee \neg P \vee Q$). Even if we restrict $R$ with consistent literal sets, this definition does not capture partial satisfaction either. For instance, $x \vee \neg y$ should partially entail $x \wedge y$ according to this definition since $(x \vee \neg y) \wedge y \models x \wedge y$. However, this conclusion is counter-intuitive. As an alternative possibility, since "$P$ partially entails $Q$" intuitively means that $P$ entails some parts of $Q$, one may define partial entailment as: a formula $P$ partially entails a formula $Q$ iff there exists $Q'$ and $Q''$ such that $\models Q \leftrightarrow Q' \wedge Q''$, $Q'' \not\models Q'$ and $P \models Q'$. However, this definition fails to capture the essence of partial entailment either. For instance, according to this definition, $x$ partially entails $y$ since $\models y \leftrightarrow (x \vee y) \wedge (\neg x \vee y)$, $\neg x \vee y \not\models x \vee y$, and $x \models x \vee y$.

According to the definitions of partial entailments, $x$ (weakly, strongly) partially entails both $x \wedge y$ and $x \vee y$. One may conclude that this result indicates that conjunction and disjunction play a similar role in partial entailments. However, this is not the case. In this example, the reasons why $x$ partially entails $x \wedge y$ and why $x$ partially entails $x \vee y$ are completely different. In the former case, the reason is that $x$ is a part of $x \wedge y$, while in the latter case, the reason is that $x$ classically entails $x \vee y$. In fact, conjunction and disjunction play totally different roles in partial entailments. For instance, $x \wedge y$ partially entails $x$, while $x \vee y$ does not. More examples can be easily found.

Another evidence to illustrate the differences between conjunction and disjunction is to consider another special case of partial entailments by restricting the formulas to clauses, i.e., disjunction of literals, and assuming the background theory to be empty.

**Proposition 10** *Let $\delta$ and $\delta'$ be two non-valid clauses (i.e., disjunctions of two consistent sets of literals). The following statements are equivalent.*

1. *$\delta$ is a subset of $\delta'$ if they are represented as two sets of literals.*

2. *$\delta$ classically entails $\delta'$.*

3. *$\delta$ weakly partially entails $\delta'$.*

4. *$\delta$ partially entails $\delta'$.*

5. *$\delta$ strongly partially entails $\delta'$.*

Proposition 10 states that under this setting, classical entailment and all three kinds of partial entailments coincide. Compared Proposition 10 to Definition 1, disjunction and conjunction are quite different in partial entailments.

As we mentioned earlier, the Substitution principle does not hold for all three kinds of partial entailments in general. This means that, in partial entailments, the atoms indeed play a different role than formulas. For instance, $x$ (weakly, strongly) partially entails $x \wedge y$. However, it is not necessary the case that $P$ partially entails $P \wedge Q$. This shows a difference between partial entailments and classical entailment due to the fact that the Substitution principle holds for standard propositional logic.

## 3. Complexity Analysis

In this section, we analyze the complexity issues in relation to the extended notion of prime implicant and all three kinds of partial entailments. We assume that readers are familiar with some basic notions of computational complexity. More details can be found in the textbook by Papadimitriou (1994).

Here, we briefly recall some complexity classes. DP is a widely used complexity class, which contains all languages $L$ such that $L = L_1 \cap L_2$, where $L_1$ is in NP and $L_2$ is in coNP. The polynomial hierarchy is defined recursively as follows:

$$\Delta_0^P := \Sigma_0^P := \Pi_0^P := P,$$
$$\Delta_{i+1}^P := P^{\Sigma_i^P}, i \geq 0,$$
$$\Sigma_{i+1}^P := NP^{\Sigma_i^P}, i \geq 0,$$
$$\Pi_{i+1}^P := coNP^{\Sigma_i^P}, i \geq 0.$$

For instance, $\Delta_3^P = P^{\Sigma_2^P}$ is the complexity class of all languages that are recognizable in polynomial time by a deterministic Turing Machine equipped with a $\Sigma_2^P$ oracle. In particular, $\Delta_3^P[O(log\, n)]$ is a subset of the $\Delta_3^P$ by restricting the uses of the oracle within logarithmical times. Gottlob (1995) showed that the complexity classes $\Delta_3^P$ and $\Delta_3^P[O(log\, n)]$ coincide with so-called $\Sigma_2^P$-dag and $\Sigma_2^P$-tree respectively. Intuitively, a $\Sigma_2^P$-dag is an acyclic directed graph of dependent queries to $\Sigma_2^P$ oracles. The nodes are queries whilst the edges represent the dependency relationships among nodes. A $\Sigma_2^P$-tree is a $\Sigma_2^P$-dag that is a tree.

We mainly investigate the following decision problems, both in the most general case and in some restricted cases:

$\texttt{PRIC}(\Gamma, \texttt{P}, \pi)$ : (Prime Implicant Checking) to determine whether a literal set $\pi$ is a prime implicant of a formula $P$ *w.r.t.* a formula set $\Gamma$.

$\texttt{LEPR}(\Gamma, \texttt{P}, \texttt{l})$ : (Literal Existence in Prime Implicants) to determine whether a literal $l$ is in at least one prime implicant of a formula $P$ *w.r.t.* a formula set $\Gamma$.

$\texttt{LAPR}(\Gamma, \texttt{P}, \texttt{l})$ : (Literal in All Prime Implicants) to determine whether a literal $l$ is in all prime implicants of a formula $P$ *w.r.t.* a formula set $\Gamma$.

$\texttt{WPE}(\Gamma, \texttt{P}, \texttt{Q})$ : (Weak Partial Entailment) to determine whether $P$ weakly partially entails $Q$ *w.r.t.* $\Gamma$.

$\texttt{PE}(\Gamma, \texttt{P}, \texttt{Q})$ : (Partial Entailment) to determine whether $P$ partially entails $Q$ *w.r.t.* $\Gamma$.

$\texttt{SPE}(\Gamma, \texttt{P}, \texttt{Q})$ : (Strong Partial Entailment) to determine whether $P$ strongly partially entails $Q$ *w.r.t.* $\Gamma$.

For convenience, we omit $\Gamma$ when it is empty.

The first three decision problems are concerned with the relativized notion of prime implicant, while the others are concerned with the notions of partial entailments. The main focuses in this paper are of course the latter ones, while the former ones are needed as the intermediate steps.

Note that some of the complexity results related to prime implicant can be found in or followed by existing results in the literature (see more details in Appendix). For instance, as shown by Marquis (2000) (Propositions 3.27 and 3.32), checking whether a literal set (with or without background theory) is a prime implicate of a formula is DP complete. Another result comes from Proposition 10 by Lang et al. (2003) that checking whether a literal is in at least one prime implicant of a formula is NP complete. Also, as a consequence of the correspondence between prime implicant and abduction, many complexity results related to abduction can be borrowed for studying prime implicant. For instance, as shown in the Appendix, checking whether a literal $l$ is in one of the prime implicants of $F$ with respect to $\Gamma$ is exactly the task of relevance checking in abduction.

However, as the main focus of this paper, partial entailments are defined upon prime implicant (i.e. abduction). Although existing results conclude some of the computational complexity results related to prime implicant, they do not reveal too much information about the complexities of partial entailments.

## 3.1 Empty Background Theory

We start our complexity analysis when the background formula set $\Gamma$ is empty. The following proposition shows the complexity results in relation to prime implicant when the background theory is empty. As mentioned above, the first two results follow by existing results in the literature.

**Proposition 11** *The computational complexities in relation to prime implicant with empty background theory are summarized in Table 3.*

Table 3: Complexity results for prime implicant: empty background theory

| | |
|---|---|
| `PRIC(P,`$\pi$`)` | DP complete |
| `LEPR(P,l)` | NP complete |
| `LAPR(P,l)` | DP complete |

Based on the computational analysis on prime implicant, the following proposition presents a complete result for the complexity results in relation to all three kinds of partial entailments when the background theory is empty.[3]

**Proposition 12** *The computational complexities in relation to partial entailments with empty background theory are summarized in Table 4. Here, $l$ is a literal, $\pi$ is a set of literals and $P$ and $Q$ are formulas.*

Table 4: Complexity results for partial entailments: empty background theory

| - | WPE | PE | SPE |
|---|---|---|---|
| $(l, P)$ | NP complete | NP complete | NP complete |
| $(P, l)$ | DP complete | DP complete | coNP complete |
| $(\pi, P)$ | NP complete | NP complete | $\Sigma_2^P$ complete |
| $(P, \pi)$ | DP complete | DP complete | coNP complete |
| $(P, Q)$ | $\Pi_2^P$ complete | $\Pi_2^P$ complete | $\Pi_3^P$ complete |

Let us take a closer look at Table 4. First of all, it shows that it is not an easy task to compute the notions of partial entailments, even when the background theory is empty. For instance, to check whether a formula (weakly) partially entails another formula is at the second level of polynomial hierarchy. More surprisingly, this task for strong partial entailment is even more difficult, which is at the third level of polynomial hierarchy.

Also, from Table 4, one can conclude that the computational complexities increase when the forms of formulas become more complicated. For instance, checking whether a literal strongly partially entails a formula is NP complete. When extending the literal to a literal set, this becomes $\Sigma_2^P$ complete. The task further turns into $\Pi_3^P$ complete when extending the antecedent to an arbitrary formula. Interestingly, for weak partial entailment and partial entailment, no matter considering partially entailing a formula or partially entailed by a formula, turning a literal to a literal set does not increase the complexity. However, for strong partial entailment, this is the case for partially entailed by a formula but not for partially entailing a formula.

---

3. Clearly, if both the antecedent and the consequent are restricted to literal sets, then the corresponding decision problems in relation to all three kinds of partial entailments can be solved in linear time.

Finally, it can be observed that, when the background theory is empty, weak partial entailment and partial entailment have the same computational complexities, while strong partial entailment acts differently. Interestingly, based on some basic assumptions in the complexity theory, checking strong partial entailment relationship is sometimes easier than the rest two (e.g. compare $\mathtt{SPE}(\mathtt{P},\pi)$ and $\mathtt{PE}(\mathtt{P},\pi)$), but sometimes more difficult (e.g. compare $\mathtt{SPE}(\mathtt{P},\mathtt{Q})$ and $\mathtt{PE}(\mathtt{P},\mathtt{Q})$).

In some application scenarios, the background theory is simply a set of facts, which can be represented as a set of literals. Hence, we are interested in the computational complexities for this special case. As we shall see later, this is much simpler than the general case when the background theory is an arbitrary set of formulas.

The following proposition indicates that, fortunately, turning the background formula set into a literal set does not increase the computational complexities for all three kinds of partial entailments.

**Proposition 13** *All the complexity results in Tables 3 and 4 remain the same if the background theory is a set of literals.*

### 3.2 General Background Theory

Finally, we face to the general case, in which the background theory is an arbitrary formula set. We only consider the general cases of all the six decision problems. One reason is that the background theory is a general setting already. Another reason is that, even for a single literal, there might exist many different prime implicants of it with respect to an arbitrary background theory.

**Proposition 14** *The computational complexities in relation to partial entailments in general case are summarized in Table 5.*[4]

Table 5: Complexity results: general case

| - | empty $\Gamma$ | arbitrary $\Gamma$ |
|---|---|---|
| $\mathtt{PRIC}(\Gamma,\mathtt{P},\pi)$ | DP complete | DP complete |
| $\mathtt{LEPR}(\Gamma,\mathtt{P},\mathtt{l})$ | NP complete | $\Sigma_2^P$ complete |
| $\mathtt{LAPR}(\Gamma,\mathtt{P},\mathtt{l})$ | DP complete | $\Pi_2^P$ complete |
| $\mathtt{WPE}(\Gamma,\mathtt{P},\mathtt{Q})$ | $\Pi_2^P$ complete | in $\Delta_3^P[O(log\,n)]/\ \Pi_2^P$ hard$/\ \Sigma_2^P$ hard |
| $\mathtt{PE}(\Gamma,\mathtt{P},\mathtt{Q})$ | $\Pi_2^P$ complete | in $\Pi_3^P/\ \Pi_2^P$ hard$/\ \Sigma_2^P$ hard |
| $\mathtt{SPE}(\Gamma,\mathtt{P},\mathtt{Q})$ | $\Pi_3^P$ complete | $\Pi_3^P$ complete |

According to Table 5, the complexities of prime implicant checking and strong partial entailment checking remain the same when allowing arbitrary background theory, while the complexities increase for the other decision problems. For instance, checking whether a literal is in one (or all) prime implicant(s) of a formula increases from NP complete (DP complete, resp.) to $\Sigma_2^P$ complete ($\Pi_2^P$ complete, resp.) when the background theory turns into an arbitrary one. Also, the complexities of checking weak partial entailment and partial entailment in general case increase a little because these two problems are both $\Sigma_2^P$ and $\Pi_2^P$ hard. However, the corresponding complexity for strong partial entailment remains the same since it is already $\Pi_3^P$ hard in the special case when the background theory is empty.

---

4. Here, we also present the corresponding computational complexity results when the background theory is empty in order to compare them directly.

As shown in Table 5, $\mathtt{WPE}(\Gamma,\mathtt{P},\mathtt{Q})$ and $\mathtt{PE}(\Gamma,\mathtt{P},\mathtt{Q})$ are both $\Pi_2^P$ hard and $\Sigma_2^P$ hard. This shows that they are neither $\Pi_2^P$ complete nor $\Sigma_2^P$ complete based on some basic assumptions in the complexity theory. We conjecture that $\mathtt{WPE}(\Gamma,\mathtt{P},\mathtt{Q})$ is $\Delta_3^P[O(log\,n)]$ complete and $\mathtt{PE}(\Gamma,\mathtt{P},\mathtt{Q})$ is $\Pi_3^P$ complete. However, to verify this, more advanced techniques, e.g. the raising technique proposed by Liberatore (2007), are needed.

## 4. Related Work

In section 2, we discussed the relationships between partial entailments and classical entailment. Here, we further consider the relationships between partial entailments and other related notions in the literature, including the family of notions of relevance and their relatives, such as formula-variable independence (Boutilier, 1994; Lang et al., 2003), relevance between formulas (Lakemeyer, 1995, 1997), novelty (Marquis, 1991) and probabilistic positive relevance (Zhou & Chen, 2006), and partial satisfaction of a formula (Lieberherr & Specker, 1981; Käppeli & Scheder, 2007).

### 4.1 Formula-Variable Independence

Lang et al. (2003) defined a notion of formula-variable independence between a formula and a set of atoms. Roughly speaking, a formula is independent to a set of atoms if the formula can be rewritten to another one that mentions no atoms from the set of atoms.

**Definition 7 (Lang et al., 2003)** *A formula $F$ is* variable-independent *to a set $V$ of variables iff there exists a formula $G$ such that $\models F \leftrightarrow G$ and $Atom(G) \cap V = \emptyset$.*

Also, Lang et al. proved that variable-independence coincides with the notions of influenceability introduced by Boutilier (1994) and relevance of a formula to a set of atoms introduced by Lakemeyer (1997).

As shown by Lang et al. (2003), the notion of formula-variable independence can be reformulated by prime implicant. Here, we show that it can be reformulated by weak partial entailment as well.

**Proposition 15** *Let $F$ be a formula and $V$ a set of atoms. The following statements are equivalent:*

1. *$F$ is variable-independent to $V$.*

2. *For all literals $l$ in $V \cup -V$, $l$ is not in any prime implicants of $F$.*

3. *For all literals $l$ in $V \cup -V$, $l$ does not weakly partially entail $F$.*

### 4.2 Other Forms of Relevance

Lakemeyer also introduced other forms of relevance as well, including strict relevance, explanatory relevance and relevance between two subject matters (Lakemeyer, 1997).

**Definition 8 (Lakemeyer, 1997)** *A formula $F$ is strictly relevant to a set $V$ of atoms iff every prime implicate of $F$ contains at least one atom from $V$.*

The following proposition shows that strict relevance can be reformulated by prime implicant and weak partial entailment as well.

**Proposition 16** *Let $F$ be a formula and $V = \{x_1, \ldots, x_n\}$ a set of atoms. The following statements are equivalent.*

1. *$F$ is strictly relevant to $V$.*

2. *$\neg F$ weakly partially entails $(x_1 \wedge \ldots \wedge x_n) \vee (\neg x_1 \wedge \ldots \wedge \neg x_n)$.*

**Definition 9 (Lakemeyer, 1997)** *A formula $F$ is explanatory relevant to a set $V$ of atoms w.r.t. a formula set $\Gamma$ iff there exists a minimal abductive explanation of $F$ w.r.t. $\Gamma$ that mentions a variable from $V$.*

The definition of explanatory relevance can be reformulated by prime implicant. That is, $F$ is explanatory relevant to $V$ w.r.t. $\Gamma$ iff there exists $l \in V \cup -V$ such that $l$ is in at least one of the prime implicants of $F$ w.r.t. $\Gamma$.

### 4.3 Relevance between Formulas

As pointed out by Lakemeyer (1997), it is interesting to consider relevance between formulas with respect to a background theory. Here, we propose a definition of relevance between two formulas with respect to a background theory by using prime implicant.

**Definition 10** *A formula $P$ is relevant to a formula $Q$ with respect to a formula set $\Gamma$ iff there exists a prime implicant $\pi$ of $P$ w.r.t. $\Gamma$ and a prime implicant $\pi'$ of $Q$ w.r.t. $\Gamma$ such that $\pi \cap \pi' \neq \emptyset$ .*

Definition 10 looks very similar to the definition of weak partial entailment (See Definition 4). The major difference between these two definitions is that weak partial entailment is defined by a $\forall \exists$ style, whilst relevance is defined by an $\exists \exists$ style. More precisely, in the former, we require that "for all" prime implicants of $P$ w.r.t. $\Gamma$, "there exists" a prime implicant of $Q$ w.r.t. $\Gamma$ such that their intersection is not empty. However, in the latter, we require that "there exists" a prime implicant of $P$ w.r.t. $\Gamma$ and "there exists" a prime implicant of $Q$ w.r.t. $\Gamma$ satisfying the same condition.

However, weak partial entailment can serve as a strict notion of relevance between two formulas with respect to a background theory. That is, a formula $P$ is *strictly relevant* to another formula $Q$ with respect to a formula set $\Gamma$ if and only if $P$ weakly partially entails $Q$ w.r.t. $\Gamma$. Clearly, this strict version of relevance between formulas w.r.t. to a background theory based on weak partial entailment implies the normal one defined in Definition 10, but the converse does not hold in general.

### 4.4 Novelty and Novelty-Based Independence

The notion of novelty (Marquis, 1991) is defined between two formulas with respect to a formula set as well.

**Definition 11 (Marquis, 1991)** *A formula $P$ is* new positive *(*new negative*) to another formula $Q$ with respect to a formula set $\Gamma$ iff there exists a prime implicant $\pi$ of $Q$ ($\neg Q$) with respect to $\Gamma \cup \{P\}$, which is not a prime implicant of $Q$ ($\neg Q$) with respect to $\Gamma$.*

In addition, a notion of independence, namely novelty-based independence (also known as separability, see Levesque, 1998), between two formulas is defined based on novelty (Lang, Liberatore, & Marquis, 2002).

**Definition 12 (Lang et al., 2002)** *Two formulas $P$ and $Q$ are* (novelty-based) indepen- dent *iff $P$ is not new negative to $Q$ w.r.t. $\top$.*

Intuitively, that $P$ is new to $Q$ w.r.t. $\Gamma$ means that adding new information $P$ to the background theory $\Gamma$ has some influences on $Q$ or $\neg Q$. Although novelty and partial entailments are both defined by using the notion of prime implicant, they are essentially different. For instance, (non)novelty-based independence satisfies Symmetry. That is, if $P$ is (non)novelty-based independent to $Q$, then $Q$ is (non)novelty-based independent to $P$. However, (weak, strong) partial entailment does not satisfy Symmetry. For instance, $x$ (weakly, strongly) partially satisfies $x \vee y$, but the converse does not hold. As another example, $x \vee y$ (weakly, strongly) partially entails $x \wedge y$. However, $x \vee y$ is not new positive (negative) to $x \wedge y$. Also, $x \leftrightarrow y$ is new positive (negative) to $x$. However, $x \leftrightarrow y$ does not (weakly, strongly) partially entail $x$.

### 4.5 Probabilistic Positive Relevance

Another approach to formalize a certain kind of usefulness is probabilistic positive relevance (Zhou & Chen, 2006), which is based on probability distributions. The basic idea of pos- itive relevance is: a formula $P$ is *positive relevant* to another formula $Q$ with respect to a background formula set $\Gamma$ iff for all probability distributions $Pr$, $Pr(Q|\Gamma \cup \{P\}) \geq Pr(Q|\Gamma)$.

Although positive relevance looks similar to partial entailments, and probability distri- bution is highly related to prime implicant (Lang et al., 2002), the underlying intuition and the semantic properties of positive relevance and partial entailments are quite different. For instance, according to the definition of positive relevance by Zhou and Chen (2006), $x \vee y$ is positive relevant to $x$. However, $x \vee y$ does not (weakly, strongly) partially entail $x$. Generally speaking, in positive relevance (defined based on probability distributions), Symmetry holds. That is, $P$ is positive relevant to $Q$ w.r.t. $\Gamma$ if and only if $Q$ is positive relevant to $P$ w.r.t. $\Gamma$. However, this does not hold for any kinds of partial entailments.

### 4.6 Partial Satisfaction of a CNF

The term "partial satisfaction" is also used for satisfaction of a subset of a set of clauses (Lieberherr & Specker, 1981; Käppeli & Scheder, 2007). A CNF formula (i.e. a conjunction of clauses) is said to be *k-satisfiable* if every subformula of it containing at most $k$ clauses is satisfiable.

Although the term "partial satisfaction" is used for both $k$-satisfaction and partial entailments, it represents different intuitive meanings. For the former, 'part" means a subset of a formula (i.e. the set of clauses), while for the latter, it means a subset of a prime implicant of a formula. Hence, partial entailments and $k$-satisfaction are basically irrelevant. Moreover, $k$-satisfaction is only concerned with a particular (CNF) formula, while partial entailments are concerned with the relationship between two formulas ($w.r.t.$ a background theory).

## 5. Partial Goal Satisfaction

In traditional logic based approach of rational agency, the agents always try to find actions to "completely" achieve their goals. If not successful, the agents would choose to wait and do nothing. This idea is usually formalized by using classical entailment. However, it is not always the case that the agents can find out such a perfect action (think about the real world that we live in). Under the situation, it is sometimes useful for the agents to do something "towards" their goals rather than just waiting, in which those actions "partially" satisfying the agents' goal should be rational candidates. Here, partial entailments may serve as a logic foundation.

**Example 6** *Let us consider an example. Suppose that Laura wants to have milk and cereal as her breakfast. However, there are only three choices available on the breakfast menu.*

**Choice1** *Only milk is offered.*

**Choice2** *Milk and bread are offered.*

**Choice3** *Only bread is offered.*

*Due to the current information, none of these choices completely satisfies Laura's goal. Then, what should she do?* □

Here, we argue that, under this circumstance, it is also rational for the agents to choose those actions partially achieving their goal according to their belief, as there is no action completely achieving the goal.

It is important to allow those actions partially achieving the agents' goal to be rational candidates. First of all, the rationality of performing those actions is solid since they make the goal closer for the agents and there is nothing better to do. Hence, choosing those actions useful to the agents' goal is more reasonable than just waiting. Also, the environment is dynamic in nature. In some cases, if the agents do not choose those actions partially achieving their goal, it might lose the chance to achieve the goal forever. This frequently happens in everyday life when we cannot seize the opportunities.

We formalize partial achievement between actions and goals under the agents' belief by using the notion of partial entailments proposed in Section 2. This is natural since partial entailments precisely capture the partial satisfaction relations between two formulas with respect to a background theory. To apply partial entailments to formalize partial goal satisfaction, we treat the background theory as the agents' belief, the consequent as the agents' goal and the antecedent as the consequences of the action. One obstacle is how to represent actions and their consequences. Here, to address this issue, we simply use triples in Hoare logic to represent actions.[5]

Again, we restrict our discussions within a propositional language. The goal and the belief of agents are represented as a propositional formula and a set of propositional formulas respectively. An *action* $\alpha$ is a triple $\langle Pre(\alpha), \alpha, Post(\alpha) \rangle$, where $\alpha$ is a label called the *body* of the action, and $Pre(\alpha)$ and $Post(\alpha)$ are two propositional formulas called the *precondition* and *postcondition* of the action respectively. Next, we show how the traditional idea of complete goal satisfaction and our idea of partial goal satisfaction can be formalized under this setting.

**Definition 13 (Complete goal satisfaction)** *Let $\Gamma$ be an agent's belief and $G$ the agent's goal. Let $\mathcal{A}$ be a set of candidate actions. An action $\alpha \in \mathcal{A}$ completely achieves the agent's goal $G$ according to its belief $\Gamma$ iff:*

1. *$\Gamma \models Pre(\alpha)$.*

2. *$\Gamma \cup \{Post(\alpha)\}$ is consistent.*

3. *$\Gamma \models Post(\alpha) \rightarrow G$.*

Definition 13 means that if the agent's belief satisfies the precondition of an action and it believes that the postcondition of the action entails its goal, then this action completely achieves its goal under the belief. Partial goal satisfaction is formalized in a similar way except that partial entailments are used instead of classical entailment.

**Definition 14 (Partial goal satisfaction)** *Let $\Gamma$ be an agent's belief and $G$ the agent's goal. Let $\mathcal{A}$ be a set of candidate actions. An action $\alpha \in \mathcal{A}$ (weakly, strongly) partially achieves the agent's goal $G$ according to its belief $\Gamma$ iff:*

1. *$\Gamma \models Pre(\alpha)$.*

2. *$Post(\alpha)(\succ_\Gamma^W, \succ_\Gamma^S) \succ_\Gamma G$.*

Definition 14 means that if the agent's belief satisfies the precondition of an action and it believes that the postcondition of the action partially entails its goal, then this action partially achieves its goal under the belief.

**Example 7** *Consider Example 6 proposed previously. To formalize this example, we use $x$, $y$ and $z$ to represent "having milk", "having cereal" and "having bread" respectively. Then, Laura's belief is empty since there is no background knowledge; her goal can be represented as $x \wedge y$; the preconditions of the three actions to take the three choices respectively are empty as well and can be represented as $\top$, while the postcondition of the three cases can be represented as $x$, $x \wedge z$ and $z$ respectively. According to Definition 13, none of the three actions completely achieves the goal $x \wedge y$. However, according to Definition 14, to take Choice1 and Choice2 (weakly) partially achieve the goal. In particular, to take Choice1 strongly partially achieves the goal, whilst to take Choice3 does not partially achieve the goal in any sense.* □

---

5. Here, we only use a simple action theory to demonstrate how the notion of partial entailments can be used for formalizing partial goal satisfaction. The problems of how to represent actions for agents in general and how to deal with the frame problem is out of the scope of this paper.

A problem arises which kind of partial goal satisfaction is rational for the agents. At first glance, it seems that partial entailment is more appropriate for this purpose. Weak partial entailment may contain side effects that are not acceptable, whilst strong partial entailment are too strict that some powerful actions are excluded. However, we believe it is better to leave this to the agent designers. One may choose different kinds of partial goal satisfaction in different application domains, e.g., strong partial entailment if both irrelevancies and side effects are crucial issues, or partial entailment if side effects are unacceptable but irrelevancies are not, or weak partial entailment in looser cases.

It is worth mentioning that partial goal satisfaction only proposes one possible solution for dealing with the situations that the agents cannot find a plan to completely achieve their goal. It is neither the only solution nor necessarily perfectly rational. In some such situations, the actions achieving some parts of the agents' goal may lose its chances to achieve other parts, which might be unacceptable for the agents.

Also, the actions partially achieving the agents' goal are only rational candidates but not compulsory. It is quite different between the fact that an action is a rational candidate and that the agents really choose to perform it. Partial goal satisfaction only explains the rationality of those actions partially achieving the goals. The problem of which particular action should be chosen to perform among all the actions partially achieving the agents' goal is another research topic and is beyond the scope of this paper.

Another approach for handling these situations is to explicitly represent the agents' full preferences among all combinations of goals. For instance, in Example 6, Laura may have the following full preferences among all candidate goals:

$$\{Milk, Cereal\} > \{Milk\} > \{Cereal\} > \{\},$$

where $>$ represents the preference relation. If Laura's original goal (i.e. $Milk \wedge Cereal$) cannot be achieved, then she intends to achieve the second best candidate goal (i.e. $Milk$) according to her preference. It can be observed that partial goal satisfaction is not always consistent with the full preference approach, for instance, if the above preference of Laura turns into $\{Milk, Cereal\} > \{\} > \{Milk\} > \{Cereal\}$. Although sharing some similarities, these two approaches are essentially different. First of all, full preference requires additional information, while partial goal satisfaction does not. Moreover, it is expensive to represent the full preference of agents. Since the number of possible combinations of goals are exponential with respect to the size of goals, to represent a full preference requires double exponential number of new information. Last but not least, it is usually difficult to obtain the full preferences of agents. Hence, we believe that both the full preference approach and the partial goal satisfaction approach have their own advantages and disadvantages for dealing with the situations when there is no perfect actions to achieve the agents' goal.

The idea of partial satisfaction of goals has been discussed elsewhere in the AI literature (Haddawy & Hanks, 1992; Smith, 2004; Do, Benton, van den Briel, & Kambhampati, 2007). One approach is so called partial satisfaction planning (Smith, 2004; Do et al., 2007). In partial satisfaction planning, agents find plans, instead of completely achieving their goals, that partially achieve their goals. Since the agents' goals in AI planning, particularly in STRIPS-like planning, are usually represented as conjunctions of small pieces of subgoals, it can be formalized as finding plans that achieve subsets of those subgoals. In fact, this is a special case of partial entailments in the sense that partial entailment is dealing with arbitrary propositional formulas. Another approach is due to Haddawy and Hanks (1992). In their approach, the agents' goals are represented as groups of candidate goals and each of them is associated with a real number between $[0,1]$ to represent the degree of chances of satisfying a goal. A plan satisfying a candidate goal with number 1 is full satisfaction while a plan satisfying a candidate goal with a number between $(0,1)$ (e.g., 0.5) is partial satisfaction. There are several differences between this approach and partial entailments. Firstly, in partial entailments, the objective formula is represented by a single formula instead of a set of candidate variations. Secondly, numerical degree of satisfaction is not introduced in partial entailments. Finally, in partial entailments, the partial satisfaction relationships come from the internal structures of formulas.

Finally, we would like to mention that the application scenario discussed here for partial entailments is reasoning about rational agents but not classical AI planning, e.g. STRIPS. The main reason is that the agents' goal and beliefs are formalized by literal sets rather than arbitrary formulas in STRIPS planning, but the main focus of this paper is to introduce the notions of partial entailments to propositional logic in general. Nevertheless, since checking partial entailments when restricted the antecedent, the consequent and the background theory all to be literal sets can be done in linear time, it is interesting to apply partial entailments in AI planning. We would like to leave this as our future investigations.

## 6. Concluding Remarks

In this paper, we introduced a new logical notion–partial entailment–to propositional logic. We distinguished three different kinds of partial entailments (See Table 1) based on the notion of prime implicant. From the weakest to strongest, they are weak partial entailment, partial entailment and strong partial entailment respectively. We then investigated the semantic properties of partial entailments (See Table 2). The results demonstrate that the properties of partial entailments are difficult to be captured since many simple inference rules do not hold. We also investigated the computational complexity of partial entailments (See Tables 4 and 5). The complexity results are surprising and interesting. For instance, checking strong partial entailment is $\Pi_3^P$ complete, even when the background theory is empty. This indicates that although the definitions of partial entailments look simple, it is not easy to compute them.

We showed that the notions of partial entailments can serve as a foundation of formalizing partial goal satisfaction in reasoning about rational agents. Another potential application scenario, mentioned in our previous work (Zhou, van der Torre, & Zhang, 2008), is goal weakening by using strong partial entailment. When an agent needs to modify its goal, it can choose a weakened one that strongly partially entails the original goal with respect to the agent's belief. This is because strong partial entailment preserves some of the important parts of the original goal. However, more sophisticated work are needed to develop such a goal weakening framework. Also, as mentioned previously, it is interesting to apply partial entailments in AI planning, where all formulas are syntactically restricted so that the computation task of partial entailment may become easier.

Another direction of future work lies in computing partial entailments. Of course, an important task is to develop an algorithm directly for this purpose. However, an alternative approach is to identify some tractable subclasses for checking partial entailments since the general complexities are relatively high.

## Acknowledgments

Some preliminary results of this paper were published (Zhou & Chen, 2004; Zhou et al., 2008). We are grateful to Xiaoping Chen and Leon van der Torre for their inspirations and contributions on these works. We are also grateful to Jerome Lang for his valuable comments on an earlier draft of this paper, and we would like to thank the anonymous reviewers for their valuable comments as well. The authors were partially supported by an Australian Research Council (ARC) Discovery Projects grant (DP0988396).

## Appendix A. Selected Proofs[6]

**Proposition 8 (Extension of Classical Entailment)** *Let $\Gamma$ be a formula set and $P$ and $Q$ two formulas nontrivial w.r.t. $\Gamma$. If $\Gamma \models P \to Q$, then $P \succ_\Gamma Q$. Also, $P \succ_\Gamma^W Q$.*

**Proof:** We first prove that $P \succ_\Gamma Q$. Let $\pi$ be a prime implicant of $P$ *w.r.t.* $\Gamma$. We have that $\Gamma \cup \pi \models P$. Since $\Gamma \models P \to Q$, we have that $\Gamma \cup \pi \models Q$. By Proposition 2, there is a subset $\pi'$ of $\pi$ such that $\pi'$ is a prime implicant of $Q$ *w.r.t.* $\Gamma$. Thus $\pi' \subseteq \pi$. Due to non-triviality of $P$ and $Q$, $\pi \cap \pi' \neq \emptyset$ and $\pi \cap -\pi' = \emptyset$. This shows that $P \succ_\Gamma Q$.

According to Proposition 6, $P \succ_\Gamma^W Q$. $\square$

**Proposition 9** *Table 2 summarizes whether all three kinds of partial entailments satisfy the inference rules considered above.*

**Proof:** Here, we only give the proofs or counterexamples of some of the results.

For Relevancy, according to Definition 5, there exists $\pi \in PI(P)$ and $\pi' \in PI(Q)$ such that $\pi \cap \pi' \neq \emptyset$. Therefore, $Atom(\pi) \cap Atom(\pi') \neq \emptyset$. It follows that $Atom(P) \cap Atom(Q) \neq \emptyset$ since $Atom(\pi) \subseteq Atom(P)$ and $Atom(\pi') \subseteq Atom(Q)$. Similar for weak partial entailment and strong partial entailment.

For Transitivity of strong partial entailment, let $\pi$ be a prime implicant of $P$ *w.r.t.* $\Gamma$. Since $P \succ_\Gamma^S Q$, there exists a literal set $\pi_1$ consistent with $\Gamma$ such that $\pi_1 \in PI(\Gamma, Q)$ and $\pi \subseteq \pi_1$. Moreover, since $Q \succ_\Gamma^S R$, there exists a literal set $\pi_2$ consistent with $\Gamma$ such that $\pi_2 \in PI(\Gamma, R)$ and $\pi_1 \subseteq \pi_2$. Hence, $\pi \subseteq \pi_2$. This shows that $P \succ_\Gamma^S R$.

Note that Transitivity does not hold for either partial entailment or weak partial entailment. For instance, $x \succ x \land y$ and $x \land y \succ y$, but $x \not\succ y$. Similarly, $x \succ^W x \land y$ and $x \land y \succ^W y$, but $x \not\succ^W y$. The reason why partial entailment and weak partial entailment fail transitivity is that they allow irrelevancies. In the above example, although $x \land y$ partially entails $y$, it contains $x$, which is irrelevant to $y$, and it is exactly the reason why $x$ partially entails $x \land y$.

---

For Left Strengthening of weak partial entailment, let $\pi$ be a prime implicant of $P$ *w.r.t.* $\Gamma$. Then, $\Gamma \cup \pi \models P$. It follows that $\Gamma \cup \pi \models R$ since $\Gamma \models P \rightarrow R$. By Proposition 2, there is a subset $\pi_1$ of $\pi$, which is a prime implicant of $R$ *w.r.t.* $\Gamma$. Moreover, $R \succ_\Gamma^W Q$. Then, there exists a prime implicant $\pi_2$ of $Q$ *w.r.t.* $\Gamma$ such that $\pi_1 \cap \pi_2 \neq \emptyset$. Thus, $\pi \cap \pi_2 \neq \emptyset$. This shows that $P \succ_\Gamma^W Q$.

Note that Left Strengthening does not hold for either partial entailment or strong partial entailment. For example, we have that $x \succ^S x \wedge y$ and $x \succ x \wedge y$, but $x \wedge \neg y \not\succ^S x \wedge y$ and $x \wedge \neg y \not\succ x \wedge y$. The reason why strong partial entailment and partial entailment fail Left Strengthening is that they prohibit side effects since the strengthening part on the left side could be side effect to the right side, e.g., $\neg y$ in the above example.

Both Left Negation and Right Negation do not hold for all three kinds of partial entailments. As an example, both $x$ and $\neg x$ (weakly, strongly) partially entail $x \leftrightarrow y$. Meanwhile, $x$ (weakly, strongly) partially entails both $x \leftrightarrow y$ and $\neg(x \leftrightarrow y)$. $\square$

**Proposition 11** *The computational complexities in relation to prime implicant are summarized in Table 3.*

**Proof:** The first result can be proved similarly to Proposition 3.27 in the work of Marquis (2000), and the second one follows directly from Proposition 10 in the work of Lang et al. (2003). For the last result, it is easy to prove that $l$ occurs in all prime implicants of $P$ iff $P$ can be satisfied and $\models P \rightarrow l$. It immediately proves the membership of this assertion. Hardness follows from the fact that $P$ is satisfiable and $Q$ is unsatisfiable iff $x$ is in all prime implicants of $(x \wedge P) \vee (\neg x \wedge Q)$, where $x$ is a new atom. $\square$

**Proposition 12** *The computational complexities in relation to partial entailments with empty background theory are summarized in Table 4. Here, $l$ is a literal, $\Pi$ is a set of literals and $P$ and $Q$ are formulas.*

**Proof:** For the membership of $\mathtt{WPE}(\pi, \mathtt{P})$, let $\pi = \{l_1, ..., l_k\}$. Then $\pi$ weakly partially entails $P$ if and only if there exists $l_i, 1 \leq i \leq k$ such that $l_i$ is in one of the prime implicant of $P$. By Proposition 11, this problem is in NP.

For the membership of $\mathtt{PE}(\pi, \mathtt{P})$, we first prove that a literal set $\pi$ partially entails a formula $P$ iff there is an assignment $\pi_1$ over $Atom \backslash Atom(\pi)$ and an assignment $\pi_2$ over $Atom(\pi)$ such that $\pi \cup \pi_1 \models P$ and $\pi_1 \cup \pi_2 \models \neg P$. "$\Rightarrow$:" By Definition 5, there is a prime implicant $\pi'$ of $P$ such that $\pi \cap \pi' \neq \emptyset$ and $\pi \cap -\pi' = \emptyset$. Let $l \in \pi' \cap \pi$. Then $\pi' \backslash \{l\} \cup \{-l\} \not\models P$. It can be extended to an assignment $\pi_0$ over $Atom$, which satisfies $\neg P$. Let $\pi_1 \subseteq \pi_0$ and $Atom(\pi_1) = Atom \backslash Atom(\pi)$; let $\pi_2 \subseteq \pi_0$ and $Atom(\pi_2) = Atom(\pi)$. Clearly, $\pi \cup \pi_1 \models P$ and $\pi_1 \cup \pi_2 \models \neg P$. "$\Leftarrow$:" By Proposition 2, there is a prime implicant $\pi'$ of $P$ such that $\pi' \subseteq \pi \cup \pi_1$. It follows that $\pi' \cap \pi \neq \emptyset$ and $\pi' \cap -\pi = \emptyset$. Hence, $\pi$ partially entails $P$. Hence, the following algorithm determines whether $\pi$ partially entails $P$: 1. simultaneously guess two literal sets $\pi_1$ and $\pi_2$; 2. check whether $\pi_1$ and $\pi_2$ together with $\pi$ satisfy the above conditions. Step 2 can be done in polynomial time. Therefore, $\mathtt{PE}(\pi, \mathtt{P})$ in NP.

For $\mathtt{SPE}(\pi, \mathtt{P})$, membership is easily shown by the following algorithm: 1. guess a consistent literal set $\pi'$; 2. check whether $\pi'$ is a prime implicant of $P$; 3. if yes, check whether $\pi$ is a subset of $\pi'$. By Proposition 11, step 2 requires an $NP$ oracle. Hence, this problem is in $\Sigma_2^P$. For hardness, we construct a reduction from $2 - QBF$. Let $X$ and $Y$ be two disjoint sets of atoms and $P$ a formula such that $Atom(P) \subseteq X \cup Y$. Let $Y = \{y_1, y_2, ..., y_k\}$; $T_1$ be $y_1 \vee ... \vee y_k$; $T_2$ be $\neg y_1 \vee ... \vee \neg y_k$; $x$ and $y$ be two new atoms different with $X \cup Y$; $Q$ be $(x \wedge y \wedge P) \vee (x \wedge \neg y \wedge T_1) \vee (\neg x \wedge y \wedge T_2)$. We now prove that $\exists X \forall Y P$ holds if and only if $x \wedge y$ strongly partially entails $Q$. Suppose that $\exists X \forall Y P$ holds. Then, there exists an assignment $\pi_0$ over $X$ such that $\pi_0 \models P$. By Proposition 2, there exists $\pi_1 \subseteq \pi_0$ such that $\pi_1$ is a prime implicant of $P$. Therefore, $\{x, y\} \cup \pi_1 \models Q$. In addition, $\{x\} \cup \pi_1 \not\models Q$. Otherwise, $\{x\} \cup \pi_1 \models (x \wedge y \wedge P) \vee (x \wedge \neg y \wedge T_1) \vee (\neg x \wedge y \wedge T_2)$. It follows that $\{x\} \cup \pi_1 \models y \vee T_1$, a contradiction. Symmetrically, $\{y\} \cup \pi_1 \not\models Q$. Moreover, for all $l \in \pi_1$, $\{x, y\} \cup \pi_1 \backslash \{l\} \not\models Q$ since $\pi_1$ is a prime implicant of $P$. This shows that $\{x, y\} \cup \pi_1$ is a prime implicant of $Q$. Hence, $x \wedge y$ strongly partially entails $Q$. On the other hand, suppose that $x \wedge y$ strongly partially entails $Q$. Then, there exists a prime implicant of $Q$ including both $x$ and $y$. Let it be $\{x, y\} \cup \pi_1$. Therefore $\pi_1 \models P$. We have $\pi_1 \not\models T_1$. Otherwise, $\{x\} \cup \pi_1 \models Q$. This shows that $\{x, y\} \cup \pi_1$ is not a prime implicant of $Q$, a contradiction. Symmetrically, $\pi_1 \not\models T_2$. This shows that $Atom(\pi_1) \subseteq X$. Moreover, $\{x, y\} \cup \pi_1 \models Q$. It follows that $\{x, y\} \cup \pi_1 \models x \wedge y \wedge P$. Therefore $\pi_1 \models P$. Let $\pi_0$ be an assignment over $X$ such that $\pi_1 \subseteq \pi_0$. Then, $\pi_0 \models P$. This shows that $\exists X \forall Y P$ holds.

For membership of $\mathtt{WPE}(\mathtt{P}, \pi)$, $P$ weakly partially entails $\pi$ iff a) $P$ is satisfiable, and b) for all assignments $\pi_1$ over $Atom \backslash Atom(\pi)$, $\pi_1 \cup -\pi \not\models P$, which is equivalent to $-\pi \models \neg P$. To prove this, suppose that $P$ weakly partially entails $\pi$. Then, $P$ is satisfiable. Now assume that there exists $\pi_1$ such that $\pi_1 \cup -\pi \models P$. By Proposition 2, there exists $\pi_2 \subseteq \pi_1 \cup -\pi$ such that $\pi_2$ is a prime implicant of $P$. However, $\pi_2 \cap \pi = \emptyset$, a contradiction. On the other hand, suppose that the above three conditions hold. Then, for all prime implicants $\pi_1$ of $P$, $\pi_1 \cap \pi \neq \emptyset$. Otherwise, $\pi_1 \cup -\pi$ can be extended to an assignment $\pi_2$ over $Atom$ such that $\pi_2 \models P$. However, $-\pi \subseteq \pi_2$, a contradiction. Hence, $\mathtt{WPE}(\mathtt{P}, \pi)$ is in DP.

For membership of $\mathtt{PE}(\mathtt{P}, \pi)$, let $\pi = \{l_1, \ldots, l_n\}$. Then, $P$ partially entails $\pi$ iff for all prime implicant $\pi'$ of $P$, a) $\pi' \cap -\pi = \emptyset$, and b) $\pi' \cap \pi \neq \emptyset$ iff a) for all $i, (1 \leq i \leq n)$, $-l_i$ is not in any of the prime implicants of $P$, and b) $P$ weakly partially entails $\pi$. Hence, by Proposition 11 and the above result for weak partial entailment, this problem is in $DP$.

For membership of $\mathtt{SPE}(\mathtt{P}, \pi)$, let $\pi = \{l_1, \ldots, l_n\}$, $Atom(P) \backslash Atom(\pi) = \{x_1, \ldots, x_m\}$. $P$ strongly partially entails $\pi$ iff all the prime implicants of $P$ are subsets of $\pi$ iff for all $i, (1 \leq i \leq n)$, $-l_i$ is not in any prime implicants of $P$ and for all $j, (1 \leq j \leq m)$, $x_j$ ($\neg x_j$) is not in any prime implicants of $P$. By Proposition 11, this problem is in coNP.

For $\mathtt{WPE}(\mathtt{P}, \mathtt{Q})$, membership can be shown by the following algorithm, which determines whether $P$ does not weakly partially entail $Q$: 1. guess a consistent literal set $\pi$; 2. check whether $\pi$ is a prime implicant of $P$; 3. if yes, check whether $\pi$ does not weakly partially entail $Q$. By Proposition 11 and the above case for $\mathtt{WPE}(\pi, \mathtt{P})$, steps 2 and 3 require an $NP$ oracle. Hence, this problem is in $\Pi_2^P$. For hardness, we construct a reduction from $2 - \overline{QBF}$. Let $X$ and $Y$ be two disjoint sets of atoms and $P$ a formula such that $Atom(P) \subseteq X \cup Y$. Let $Y = \{y_1, y_2, ..., y_k\}$. Formula $T$ is $(y_1 \vee ... \vee y_k) \wedge (\neg y_1 \vee ... \vee \neg y_k)$. We prove that $\exists X \forall Y P$ holds if and only if $P$ does not weakly partially entail $T$. Suppose that $\exists X \forall Y P$ holds. Then, there exists an assignment $\pi_1$ over $X$ such that $\pi_1 \models P$. By Proposition

51

2, there exists a prime implicant $\pi_2$ of $P$ such that $\pi_2 \subseteq \pi_1$. Clearly, there is no prime implicant $\pi_3$ of $T$ such that $\pi_2 \cap \pi_3 \neq \emptyset$. On the other hand, suppose that $P$ does not weakly partially entail $T$. Then, there exists a prime implicant $\pi_1$ of $P$ such that for all prime implicants $\pi_2$ of $T$, $\pi_1 \cap \pi_2 = \emptyset$. It follows that $Atom(\pi_1) \subseteq X$. Then, $\pi_1$ can be extended to an assignment $\pi_3$ over $X$ such that $\pi_3 \models P$. Hence, $\exists X \forall Y P$ holds.

For $\texttt{PE(P,Q)}$, membership can be shown by the following algorithm, which determines whether $P$ does not partially entail $Q$: 1. guess a consistent literal set $\pi$; 2. check whether $\pi$ is a prime implicant of $P$; 3. if yes, check whether $\pi$ does not partially entail $Q$. By Proposition 11 and the above case for $\texttt{PE}(\pi,\texttt{P})$, steps 2 and 3 require an $NP$ oracle. Hence, this problem is in $\Pi_2^P$. For hardness, we construct a reduction from $2 - \overline{QBF}$. Let $X$ and $Y$ be two disjoint sets of atoms and $P$ a formula such that $Atom(P) \subseteq X \cup Y$. Let $X = \{x_1, x_2, ..., x_k\}$; $X' = \{x_1', x_2', ..., x_k'\}$ be $k$ new atoms different with $Atom$. Formula $K$ is $(x_1 \leftrightarrow x_1') \wedge (x_2 \leftrightarrow x_2') \wedge ... \wedge (x_k \leftrightarrow x_k')$. We prove that $\forall X \exists Y P$ holds if and only if $x \wedge K$ partially entails $x \wedge P$, where $x$ is a new atom. Notice that all the prime implicants of $K$ have the form $\pi \cup \pi'$, where $\pi$ ($\pi'$) is an assignment over $X$ ($X'$) and for all $i, (1 \leq i \leq k)$, $x_i \in \pi$ iff $x_i' \in \pi'$ ($\neg x_i \in \pi$ iff $\neg x_i' \in \pi'$). Now suppose that $x \wedge K$ partially entails $x \wedge P$. Then, for all assignments $\pi_0$ over $X$, $\{x\} \cup \pi_0 \cup \pi_0'$ partially entails $x \wedge P$. Therefore, there exists an assignment $\pi_1$ over $Y$ such that $\{x\} \cup \pi_0 \cup \pi_0' \cup \pi_1 \models x \wedge P$. It follows that $\pi_0 \cup \pi_1 \models P$. This shows that $\forall X \exists Y P$ holds. On the other hand, suppose that $\forall X \exists Y P$ holds. Then, for all assignments $\pi_0$ over $X$, there exists an assignment $\pi_1$ over $Y$, such that $\pi_0 \cup \pi_1 \models P$. Therefore, for all prime implicants $\{x\} \cup \pi_0 \cup \pi_0'$ of $x \wedge K$, there exists an assignment $\pi_1$ over $Y$ such that $\{x\} \cup \pi_0 \cup \pi_0' \cup \pi_1 \models x \wedge P$. Moreover, there exists an assignment $\pi_2 = \{\neg x\} \cup \pi_0 \cup \pi_0'$ over $\{x\} \cup X \cup X'$, such that $\pi_2 \cup \pi_1 \models \neg(x \wedge P)$. Hence, $x \wedge K$ partially entails $x \wedge P$.

Finally for $\texttt{SPE(P,Q)}$, membership can be shown by the following algorithm, which determines whether $P$ does not strongly partially entail $Q$: 1. guess a consistent literal set $\pi$; 2. check whether $\pi$ is a prime implicant of $P$; 3. if yes, check whether $\pi$ doesn't strongly partially entails $Q$. By Proposition 11, step 2 requires an $NP$ oracle; by the above case for $\texttt{SPE}(\pi,\texttt{P})$, step 3 requires a $\Sigma_2^P$ oracle. Hence, this problem is in $\Pi_3^P$. For hardness, we construct a reduction from $3 - \overline{QBF}$. Let $X$, $Y$ and $Z$ be three disjoint sets of atoms and $P$ a formula such that $Atom(P) \subseteq X \cup Y \cup Z$. Suppose that $X = \{x_1, x_2, ..., x_k\}$. Let $X' = \{x_1', x_2', ..., x_k'\}$ be $k$ new atoms. Let $K$ be the formula $(x_1 \leftrightarrow x_1') \wedge (x_2 \leftrightarrow x_2') \wedge ... \wedge (x_k \leftrightarrow x_k')$. Suppose that $Z = \{z_1, z_2, ..., z_k\}$. Let $T_1$ be $z_1 \vee ... \vee z_k$ and $T_2$ be $\neg z_1 \vee ... \vee \neg z_k$ respectively. Let $R$ be the formula $x \wedge y \wedge K$ and $Q$ be the formula $(x \wedge y \wedge P \wedge K) \vee (x \wedge \neg y \wedge T_1 \wedge K) \vee (\neg x \wedge y \wedge T_2 \wedge K)$ respectively, where $x$ and $y$ are two new atoms. Next, we will prove that $\forall X \exists Y \forall Z P$ holds if and only if $R$ strongly partially entails $Q$. The proof is tedious. On the one hand, suppose that $\forall X \exists Y \forall Z P$ holds. Given a prime implicant of $R$ of the form $\{x, y\} \cup \pi \cup \pi'$, where $\pi$ is an assignment over $X$ and $\pi'$ is a corresponding assignment over $X'$. Then, there exists an assignment $\pi_1$ over $Y$ such that $\pi \cup \pi_1 \models P$. Therefore, $\{x, y\} \cup \pi \cup \pi' \cup \pi_1 \models x \wedge y \wedge P \wedge K$. It follows that $\{x, y\} \cup \pi \cup \pi' \cup \pi_1 \models Q$. By Proposition 2, there exists a subset $\pi_2$ of $\{x, y\} \cup \pi \cup \pi' \cup \pi_1$, which is a prime implicant of $Q$. We have that $x \in \pi_2$. Otherwise, $\{y\} \cup \pi \cup \pi' \cup \pi_1 \models Q$. Therefore $\{y\} \cup \pi \cup \pi' \cup \pi_1 \models x \vee T_2$, a contradiction. Symmetrically, $y \in \pi_2$. Moreover, for each atom $l \in \pi \cup \pi'$, we have that $l \in \pi_2$ since $\pi_2 \models K$. Therefore $\{x, y\} \cup \pi \cup \pi' \subseteq \pi_2$. This shows that for all prime implicants of $R$, there exists a prime implicant of $Q$ such that the former is a subset of the latter. Hence, $R$ strongly partially entails $Q$. On the other hand, suppose that $R$ strongly partially entails $Q$. Notice that for all assignments $\pi$ over $X$, $\{x, y\} \cup \pi \cup \pi'$ is a prime implicant of $R$. Therefore, there is a prime implicant of $Q$ that contains $\{x, y\} \cup \pi \cup \pi'$. Let it be $\{x, y\} \cup \pi \cup \pi' \cup \pi_1$, where $Atom(\pi_1) \subseteq Y \cup Z$. Therefore, $\pi \cup \pi_1 \models P$. We have that $\pi_1 \not\models T_1$. Otherwise, $\{x\} \cup \pi \cup \pi' \cup \pi_1 \models Q$, a contradiction. Symmetrically, $\pi_1 \not\models T_2$. This shows that $Atom(\pi_1) \subseteq Y$. Thus, $\pi_1$ can be extended to an assignment $\pi_2$ over $Y$. We have that $\{x, y\} \cup \pi \cup \pi' \cup \pi_2 \models Q$. Therefore $\{x, y\} \cup \pi \cup \pi' \cup \pi_2 \models x \wedge y \wedge P \wedge K$. Hence, $\pi \cup \pi_2 \models P$. This shows that for all assignments $\pi$ over $X$, there exists an assignment $\pi_2$ over $Y$, such that $\pi \cup \pi_2 \models P$. That is, $\forall X \exists Y \forall Z P$ holds. □

**Proposition 13** *All the complexity results in Tables 3 and 4 remain the same if the background theory is a set of literals.*

**Proof:** This assertion follows directly from the following fact:

> Suppose that $P$ is a formula and $\pi$ is a consistent set of literals. Then, $PI(\pi, P) = PI(P|\pi)$.

On one hand, suppose that $\pi_1 \in PI(\pi, P)$. Then, according to the definition, $\pi_1 \cup \pi$ is consistent and $\pi_1 \cup \pi \models P$. We have that $Atom(\pi_1) \cap Atom(\pi) = \emptyset$. Otherwise, suppose that $l \in Atom(\pi_1) \cap Atom(\pi)$, then $\pi_1 \cup \pi \backslash \{l\} \models P$. Also, $\pi_1 \cup \pi \backslash \{l\}$ is consistent. This shows that $\pi_1$ is not a prime implicant of $P$ w.r.t. $\pi$, a contradiction. Moreover, $\pi_1 \cup \pi \models P$. It follows that $\pi_1 \models P|\pi$. In addition, there does not exist $\pi_2 \subset \pi_1$ such that $\pi_2 \models P|\pi$. Otherwise, $\pi_2 \cup \pi \models P$ and $\pi_2 \cup \pi$ is consistent. This shows that $\pi_1$ is not a prime implicant of $P$ w.r.t. $\pi$, a contradiction. Hence, $\pi_1$ is a prime implicant of $P|\pi$. On the other hand, suppose that $\pi_1$ is a prime implicant of $P|\pi$. Then, $Atom(\pi_1) \cap Atom(\pi)$ is empty since $Atom(P|\pi) \cap Atom(\pi)$ is empty. In addition, $\pi_1 \cup \pi \models P$ since $\pi_1 \models P|\pi$. Thus, $\pi_1 \cup \pi$ is consistent. Moreover, there does not exist $\pi_2 \subset \pi_1$ such that $\pi_2 \cup \pi \models P$. Otherwise, $\pi_2 \models P|\pi$. This shows that $\pi_1$ is not a prime implicant of $P|\pi$, a contradiction. Hence, $\pi_1$ is a prime implicant of $P$ w.r.t. $\pi$. □

**Proposition 14** *The computational complexities in relation to partial entailment in general case are summarized in Table 5.*

**Proof:** The DP completeness for $\mathtt{PRIC}(\Gamma, \mathtt{P}, \pi)$ follows directly from Proposition 1 and the DP completeness for $\mathtt{PRIC}(\mathtt{P}, \pi)$. This can also be proved in a similar way to the techniques introduced by Marquis (2000) for proving the DP completeness for the corresponding decision problem of prime implicate (see Marquis, 2000, Proposition 3.36).

For $\mathtt{LEPR}(\Gamma, \mathtt{P}, \mathtt{l})$, the membership is easy by guessing a literal set $\pi$ and checking if $l \in \pi$ and $\pi \in PI(\Gamma, P)$. For hardness, it can be shown that $\exists X \forall Y P$ iff $x$ is in one of the elements in $PI(\Gamma, F)$, where $\Gamma = \{\neg(x \wedge P \wedge (y_1 \vee \cdots \vee y_k))\}$, $F = x \wedge P \wedge \neg(y_1 \vee \cdots \vee y_k) \vee \neg x \wedge (\neg y_1 \vee \cdots \vee \neg y_k)$, where $x$ is a new atom. The proof is tedious. We only outline the basic ideas as follows.

$x$ is in one of the elements of $PI(\Gamma, F)$

iff

$\exists \pi, x \in \pi, \pi \not\models \neg \bigwedge \Gamma, \pi \models \bigwedge \Gamma \to F$ and $\forall \pi' \subset \pi, \pi' \not\models \bigwedge \Gamma \to F$.

iff

$\exists \pi_1, \pi_1 \cup \{x\} \not\models \neg \bigwedge \Gamma, \pi_1 \cup \{x\} \models \neg \bigwedge \Gamma \vee F$ and $\pi_1 \not\models \neg \bigwedge \Gamma \vee F$. (Notice that $\pi_1 \cup \{x\}$ is not necessarily the same as $\pi$ mentioned above.)

iff

$\exists \pi_1, \pi_1 \not\models (\neg \bigwedge \Gamma)|x, \pi_1 \models (\neg \bigwedge \Gamma)|x \vee F|x$ and $\pi_1 \not\models (\neg \bigwedge \Gamma)|\neg x \vee F|\neg x$.

iff

$\exists \pi_1, \pi_1 \not\models P \wedge (y_1 \vee \cdots \vee y_k), \pi_1 \models P$ and $\pi_1 \not\models \neg y_1 \vee \cdots \vee \neg y_k$.

iff

$\exists X \forall Y P$ holds. In fact, the hardness result also follows from the $\Sigma_2^P$ completeness of checking relevance in abductive reasoning (see Eiter & Gottlob, 1995, Thm. 4.11).

For `LAPR(Γ,P,l)`, the membership can be shown by the following algorithm, which determines whether $l$ is not in all prime implicants of $P$ $w.r.t.$ $\Gamma$: 1. guess a literal set $\pi$; 2. check if $\pi$ is a prime implicant of $P$ $w.r.t.$ $\Gamma$; 3. if yes, check if $l$ is not in $\pi$. According to the above result for prime implicant checking, this problem is in $\Sigma_2^P$. Thus, the original problem is in $\Pi_2^P$. For harness, we can prove it in a similar way to the above proof for the $\Sigma_2^P$ hardness of `LEPR(Γ,P,l)`. Indeed, one can prove that $\exists X \forall Y P$ iff $x$ is in one of the elements of $PI(\Gamma, F)$, where $\Gamma = \{\neg(x \wedge z \wedge P \wedge (y_1 \vee \cdots \vee y_k))\}$, $F = x \wedge z \wedge P \wedge \neg(y_1 \vee \cdots \vee y_k) \wedge \neg x \wedge \neg z \wedge (\neg y_1 \vee \cdots \vee \neg y_k)$, where $x$ and $z$ are two new atoms. Moreover, $x$ is in one of the elements of $PI(\Gamma, F)$ iff $\neg x$ is not in all of the elements in $PI(\Gamma, F)$ since every element in $PI(\Gamma, F)$ contains either $x$ or $\neg x$. This shows that determining whether a literal is not in all prime implicants of a formula $w.r.t.$ a formula set is $\Sigma_2^P$ hard. It follows that `LAPR(Γ,P,l)` is $\Pi_2^P$ hard.

For `WPE(Γ,P,Q)`, the $\Pi_2^P$ hardness follows directly from Proposition 12. The $\Sigma_2^P$ hardness can be implied from the hardness proof of the $\Sigma_2^P$ hardness of `LEPR(Γ,P,l)` by noticing that the only prime implicant of $x$ $w.r.t.$ $\Gamma$ is $\{x\}$ itself. For the membership, let us first consider the following algorithm, which determines whether $P$ does not weakly partially entail $Q$ $w.r.t.$ $\Gamma$: 1. compute all the literals which occur at least in one of the prime implicants of $Q$ $w.r.t.$ $\Gamma$; 2. check whether there is a prime implicant of $P$ $w.r.t.$ $\Gamma$, which does not contain any of the literals computed in step 1. According to above result for the $\Sigma_2^P$ hardness of `LEPR(Γ,P,l)`, step 1 requires linear calls to a $\Sigma_2^P$ oracle. In addition, step 2 requires only one recall to a $\Sigma_2^P$ oracle based on the results obtained in step 1. To do this, we just need to guess a consistent literal set, and to check if it is a prime implicant of $P$ $w.r.t.$ $\Gamma$ and contains no literals computed from step 1. This algorithm can be converted to a $\Sigma_2^P$-tree, in which the root is corresponding to the $\Sigma_2^P$ call for step 2, and its children are corresponding to the linear $\Sigma_2^P$ calls in step 1 for computing those literals. Thus, according the $\Sigma_2^P$-tree techniques introduced by Gottlob (1995), `WPE(Γ,P,Q)` is in $\Delta_3^P[O(log\ n)]$.

For `PE(Γ,P,Q)`, both the $\Pi_2^P$ hardness and the $\Sigma_2^P$ hardness can be shown in a similar way to the corresponding tasks for weak partial entailment. For membership, the following algorithm determines whether there exists a prime implicant $\pi'$ of $Q$ $w.r.t.$ $\Gamma$, such that $\pi \cap \pi' = \emptyset$ or $\pi \cap -\pi' \neq \emptyset$: 1. guess such a literal set $\pi'$; 2. check if $\pi'$ is a prime implicant of $Q$ $w.r.t.$ $\Gamma$; 3. check if $\pi$ and $\pi'$ satisfy the conditions. According to the above result for prime implicant checking, step 2 requires calls to an NP oracle. Thus, the above decision problem is in $\Sigma_2^P$. Based on this result, the following algorithm determines whether $P$ does not partially entail $Q$ *w.r.t.* $\Gamma$: 1. guess a literal set $\pi$; 2. check if $\pi$ is a prime implicant of $P$ *w.r.t.* $\Gamma$; 3. check if there exists a prime implicant $\pi'$ of $Q$ *w.r.t.* $\Gamma$ such that $\pi \cap \pi' = \emptyset$ or $\pi \cap -\pi' \neq \emptyset$. It is easy to see that this problem is in $\Sigma_3^P$. Thus, the original problem is in $\Pi_3^P$.

Finally, for $\mathtt{SPE}(\Gamma, \mathtt{P}, \mathtt{Q})$, hardness follows directly from Proposition 12, while membership can be shown in a similar way to the membership of the same task for partial entailment. □